\documentclass[aps,pra,twocolumn,showpacs,nofootinbib,longbibliography,notitlepage]{revtex4}
\usepackage{etex}
\usepackage{amsmath,amssymb,amsthm}
\usepackage[colorlinks=true,citecolor=blue,urlcolor=blue]{hyperref}
\usepackage[pdftex]{graphicx}
\usepackage{times,txfonts}
\usepackage{braket}
\usepackage{color}
\usepackage{natbib}
\usepackage{amsmath,blkarray}
\usepackage{mathtools}
\usepackage{latexsym}
\usepackage{tabularx, booktabs}
\usepackage{graphics,epstopdf}
\usepackage{graphicx}
\usepackage{float}
\usepackage{graphicx}
\usepackage{amsfonts}
\usepackage{subcaption}
\usepackage{color,soul}

\newcommand{\be}{\begin{equation}}
\newcommand{\ee}{\end{equation}}
\newcommand{\ba}{\begin{eqnarray}}\newcommand{\na}{\nonumber}
\newcommand{\ea}{\end{eqnarray}}
\newcommand{\fn}{\frac{1}{n}}

\begin{document}
	\title{Optimal quantum violations of $n$-locality inequalities with conditional dependence on inputs}
		\author{Sneha Munshi}
	\author{ A. K. Pan }
	\email{akp@phy.iith.ac.in}
	\affiliation{Department of Physics, Indian Institute of Technology Hyderabad Kandi, Telengana 502284, India}
    
	\begin{abstract}
Bell experiment in the network gives rise to a form of quantum nonlocality which is conceptually different from traditional multipartite Bell nonlocality. Conventional multipartite Bell experiment features a single source that distributes physical systems to multiple parties. In contrast, the network Bell experiment features multiple independent sources.  This work considers a nontrivial quantum network, the star-network configuration in an arbitrary input scenario involving  $n$ independent sources and $(n+1)$ parties, including $n$ edge parties and one central party. Each of the $n$ edge parties  shares a physical system with the central party. We consider that the central party received an arbitrary $m$ number of inputs, and each edge party receives   $2^{m-1}$ number of inputs. The joint probabilities  of the system  are bounded by some linear constraints. We show that this behaviour of the joint probabilities  in turn impose conditional dependence on the inputs of the edge parties such that the observables of each edge party  are bounded by few linear constraints. We derive a family of generalized $n$-locality inequalities and demonstrate its optimal quantum violation. We  introduce an elegant sum-of-squares approach that enables the optimization in quantum theory without specifying the dimension of the quantum system. The optimal quantum value self-tests the   observables of each edge party along with the conditional dependence. The  observables of the 
central party along with the quantum state are also self-tested from the optimization procedure itself.  Further, we characterize the network nonlocality and examine its correspondence with suitably derived standard Bell nonlocality.       
\end{abstract}
	\maketitle
	\section{Introduction}
	Bell's theorem \cite{Bell} is considered to be one of the most fundamental results in the foundations of quantum theory. It states that attempts of providing a more complete specification of reality than quantum theory in terms of a local ontological
model are not possible. The phenomenon is commonly known as quantum nonlocality, revealed through the quantum violation of suitably formulated Bell inequalities. Besides its immense impact on quantum foundations, Bell's theorem certifies the quantum correlation in a device-independent way. This eventuality paved the path for a variety of potential practical applications (see, for a review \cite{brunnerreview}) in quantum information science and quantum communication.

     The conventional bipartite Bell experiment depicts the scenario where two observers share a physical system originating from a single common source, and each observer randomly performs measurements on their respective subsystems. In a local model, the outcomes of the measurements are assumed to be fixed by a hidden variable $\lambda$. The outcomes of one party are independent of the outcomes and measurement settings of the other party. In turn, the bipartite  joint probability distribution  takes the following factorized form  \cite{Bell}
    \ba P(a,b|x,y,\lambda)=\int P(a|x,\lambda)  P(b|y,\lambda) \rho(\lambda) d\lambda\ea

    In quantum theory,  if the source distributes a  suitable entangled state and the two parties perform locally incompatible measurements,  the joint probability cannot always be factorized. Such a feature is widely known as quantum nonlocality \cite{brunnerreview} and commonly demonstrated through the violation of suitable Bell's inequality \cite{CHSH1969}.
	
	
	 Multipartite nonlocality is the a well studied
	  generalization of bipartite Bell nonlocality where multiple parties share a common entangled state originating from a single source. Multipartite entanglement and nonlocality have extensively been studied for the last two decades\cite{brunnerreview,Horodecki2009,Guhnea}. In contrast,  the multiparty   Bell experiment in the network features multiple independent sources. 	The simplest nontrivial tripartite Bell experiment in the network was first put forward by Branciad \emph{et.al} \cite{Bran2010,Cyril2012} which is widely known as the bilocality scenario. Such a network features two independent sources and three distant observers, where each source shares physical systems with two observers.  To reveal the nonlocality in quantum network  a nonlinear bilocal inequality was formulated \cite{Cyril2012}. It is shown \cite{Gisin2017}  that all pure entangled quantum states exhibit nonlocal correlations in the bilocal scenarios. A   well studied generalization of the bilocal network scenario is the $n$-local scenario in star-network scenario configuration \cite{Frit2012,Armi2014}  which is made up of a central party, surrounded by $n$ edge parties such that each of the  $n$ independent sources is sharing a physical system with the central party and an edge party. 
	
	 Of late, quantum network is being widely explored  through various types of network configurations  and  a  flurry of works has been reported \cite{Tavakoli2016,Tava2016,Frit2016,Rosse2016,Chav2016,Tava2017,Andr2017,Fras2018,Luo2018,Lee2018,Gupta2018,Cyril2019,Renou2019,Kerstjens2019,Aberg2020,Banerjee2020,Gisi2020,Supic2020,Tava2021,kundu2020,Luo2020,Banc2021,Roy2021, Jones2021,Kers2019,Chaves2021}. An interesting form of  quantum nonlocality in networks particularly features   the  without  inputs network scenario  which generates genuine quantum nonlocality \cite{Renou2019}.  The quantum violation of network nonlocality inequalities is often assessed relative to a limit derived from assuming only the no-signaling condition and independence of the sources \cite{Gisi2020}. However, the independence factor of the sources has been parametrized to show that an arbitrarily small level of independence can exhibit quantum nonlocality in networks \cite{Supic2020}. Quantum networks are capable of exploiting device-independent information processing  \cite{Lee2018}. The experimental test of quantum violations of various bilocality and $n$-locality inequalities in star-network configuration have been reported   \cite{Saunders2017,Andreoli2017,Carvacho2017,Sun2019,Poderini2020}.
	 
	To date, most of the works on quantum network assumes two inputs for each party. It is also important to note that the optimal quantum violations of $n$-locality inequalities are commonly derived by assuming two-qubit entangled states except for a work by us \cite{Sneha2021}. Therefore, it is then interesting to explore the generalized $n$-locality scenario with an arbitrary number of inputs and derive the optimal quantum violation of a proposed $n$-locality inequality without specifying the dimension of the system. For device-independent certification protocols, the optimal quantum value requires to be dimension-independent. In a  recent work \cite{Sneha2021}, we proposed  $n$-locality inequalities in star-network configuration where each of the edge parties receives   $m$ number inputs, and the central party receives  $2^{m-1}$ number of inputs. We showed that the optimal quantum violation requires the observables of each edge party to be mutually anti-commuting. However, if we exchange the number of inputs between edge parties and central parties, it becomes a nontrivial scenario that is not straightforward. Due to the complex structure of the inequalities, the optimization becomes really involved. In the present work, the optimal quantum violation of $n$-locality inequality requires a set of mutually commuting (anti-commuting) observables for the central party when the number of edge parties $n$ is even (odd). Hence, in the present work, the optimal quantum violation certifies a different set of observables compared to the set of observables required in \cite{Sneha2021}.     
	
	This work considers the star-network scenario that features $n$ number of independent sources, each sharing a physical system with an edge party and the central party. We generalize the $n$-locality scenario for arbitrary $m$ number inputs for the central party and  $2^{m-1}$ number of inputs for each edge party. As mentioned, the star-network configuration has been explored in literature by considering all parties receive two inputs, and the optimal quantum violation of $n$-locality inequality is derived by assuming that each edge party shares a two-qubit entangled state with the central party. We consider that the joint probabilities are constrained by some linear conditions. This in turn provides the conditional dependence relations of the inputs.  We propose a $n$-locality inequality in this arbitrary input scenario and derive its optimal quantum violation. Importantly, the dimension of the system is not assumed for optimization. We introduce a sum-of-squares (SOS) approach to derive the optimal quantum value without assuming the dimension. We further show that the optimal quantum value uniquely fixes the state and observables. Moreover we certify the conditional dependence of the obsevables of each edge party from the derivation of optimal quantum violation. 

We demonstrate that the optimal quantum value can be obtained when  each edge party shares at least $\lfloor\frac{m}{2}\rfloor$ copies of two-qubit maximally entangled states with the central party. Although our derivation is dimension independent,  for $m=2$,  the optimal quantum violation of $n$-locality inequality can be achieved when each edge party shares a  two-qubit entangled state with the central party, and the same holds for $m=3$. However, from $m>3$, we show that a higher-dimensional system is required. In other words, a single copy of a two-qubit entangled state may not violate the proposed $n$-locality inequality for $m>3$, but multiple copies of it can activate the non-$n$-locality for arbitrary $m$ in star-network configuration.

The plan of the paper is the following: In sec \ref{II}, we present the basic concept of the star-network and $n$-locality inequalities where each party performs the measurements of two dichotomic observables. In sec \ref{III}, we derive the optimal quantum violation of the  $n$-locality inequality when the central party of the star-network performs three dichotomic measurements, and each of the edge parties perform the measurements of four dichotomic observables. In sec \ref{IV}, we propose the  $n$-locality inequality for four dichotomic measurements of the central party and eight dichotomic measurements for each of the edge parties in the star-network configuration. The complete derivation of the inequalities  and it's  optimal quantum violation for $m=4$ and $m=5$ are  provided in Appendix \ref{A} and Appendix \ref{B} respectively . Then in sec \ref{V}, we propose the generalized form of $n$-locality inequality for arbitrary $m$ number of measurements of the central party and $2^{m-1}$ number of measurements of each of the edge parties. Using an elegant SOS approach, we derive optimal quantum violation and fix the required observables. In sec \ref{VI}, we  characterize the non-$n$-locality and demonstrate one-to-one  correspondence between the optimal violations of the generalized $n$-locality inequalities and suitable Bell-type inequalities. Finally, in sec \ref{VII}, we summarize our results and conclude with some relevant open questions.    
\section{preliminaries}\label{II}
Before presenting the main results, we briefly summarize the notion of $n$-nonlocality in star-network. The star-network configuration  \cite{Armi2014} features  $n$ independent sources and total $(n+1)$ parties. There are $n$ edge parties (Alice$_{k}$ with $k\in [n]$) and one central party. Each of the parties performs diachotomic measurements. Each source $S_{k}$ (with $k\in [n]$) shares physical systems with Alice$_{k}$ and Bob. In standard scenario, Alice$_k$  receives two inputs $x_{k}$ where $x_k\in\{1,2\}$ according to which she performs two binary outcome measurements and produce outputs $a_{k}\in \{0,1\}$. Bob performs two binary outcome measurements according to the inputs $i\in\{1,2\}$, on the joint system he receives from $n$ independent sources, and produce outcomes $b\in \{0,1\}$. 

The independence of the sources plays crucial role in network  scenario and constitute the assumption of $n$-locality. In a $n$-local model, we assume that the hidden variables $\lambda_{k}$s corresponding to the sources $S_{k}$ distributed according to $\rho(\lambda_{k})$s are independent to each other. Hence, the joint distribution  $\rho{(\lambda_{1},\lambda_{2},\cdots \lambda_{n} )}$ can be written in a  factorized form as  
	\ba
	\label{fac}
	\rho{(\lambda_{1},\lambda_{2},\cdots \lambda_{n})} =\prod\limits_{k=1}^n \rho_{k}{(\lambda_{k})}
	\ea 
	which is the $n$-locality condition. Here, each $\rho_{k}{(\lambda_{k})}$   satisfies the normalization condition  $\int d\lambda_{k}\rho_{k}{(\lambda_{k})}=1$. Using the $n$-locality condition for a star-network scenario, the  joint probability distribution  can be written as 
	\begin{eqnarray}\label{facn}
		&&P(a_{1}, a_{2}, \cdots , a_n, b,|x_{1},x_{2},\cdots x_n, i)\\
		\nonumber
	&=&\int\bigg(\prod\limits_{k=1}^n \rho_{k}{(\lambda_{k})}\hspace{3pt} d\lambda_{k}\hspace{3pt}  P(a_{k}|x_{k},\lambda_{k})\bigg)
	\hspace{5mm}\times P(b|i,\lambda_{1},\lambda_{2}\cdots \lambda_{n}).
	\end{eqnarray}
	 Clearly,
	Alice$_{k}$'s outcome  solely depends on $\lambda_{k}$, but Bob's outcome depends on all of the  $\lambda_{k}$s, where $k\in[n]$.

A suitable form of  $n$-locality inequality has been defined \cite{Armi2014,Sneha2021} as  \begin{equation}
		(\Delta_{2}^{n})_{nl}=|{I}^{n}_{2,1}|^{1/n}+|{I}^{n}_{2,2}|^{1/n}\leq {2}
		\end{equation} where '$nl$' denotes $n$-locality where  $I^{n}_{2,1}$ and $I^{n}_{2,2}$ is defined as 
\begin{eqnarray}
I^{n}_{2,1}&=&\langle(A^{1}_{1}+A^{1}_{2})(A^{2}_{1}+A^{2}_{2})\cdots (A^{n}_{1}+A^{n}_{2})B_{1}\rangle \\\nonumber
 I^{n}_{2,2}&=&\langle(A^{1}_{1}-A^{1}_{2})(A^{2}_{1}-A^{2}_{2})\cdots (A^{n}_{1}-A^{n}_{2})B_{2}\rangle\end{eqnarray}
 
Here, $A^k_{1}(A^k_{2})$ are  the observbles of Alice$_k$ corresponding to the input $x_k=1(2)$ and Bobs observables are denoted as $B_1$ or $B_2$ according to it's input $i=1,2$. 

The optimal quantum value for this $n$-locality inequality is \ba(\Delta_{2}^{n})_{Q}^{opt}=2\sqrt{2}\ea For optimization, each Alice$_k$ needs a pair of anti-commuting observables. For each $k\in[n]$, the source $S_k$  produces a two-qubit maximally entangled state which is shared between Alice$_k$ and Bob. Note that the optimal value was derived in \cite{Armi2014} and in other subsequent works by assuming that  two-qubit entangled state has been  shared between each Alice and Bob. In this work, we generalize the scenario for arbitrary input  and provide a dimension-independent optimal quantum violation of $n$-locality inequality.

	\section{ $n$-locality scenario in star-network for  $m=3$}\label{III}
	\begin{figure}[h]\begin{center}
				\includegraphics[scale=0.4]{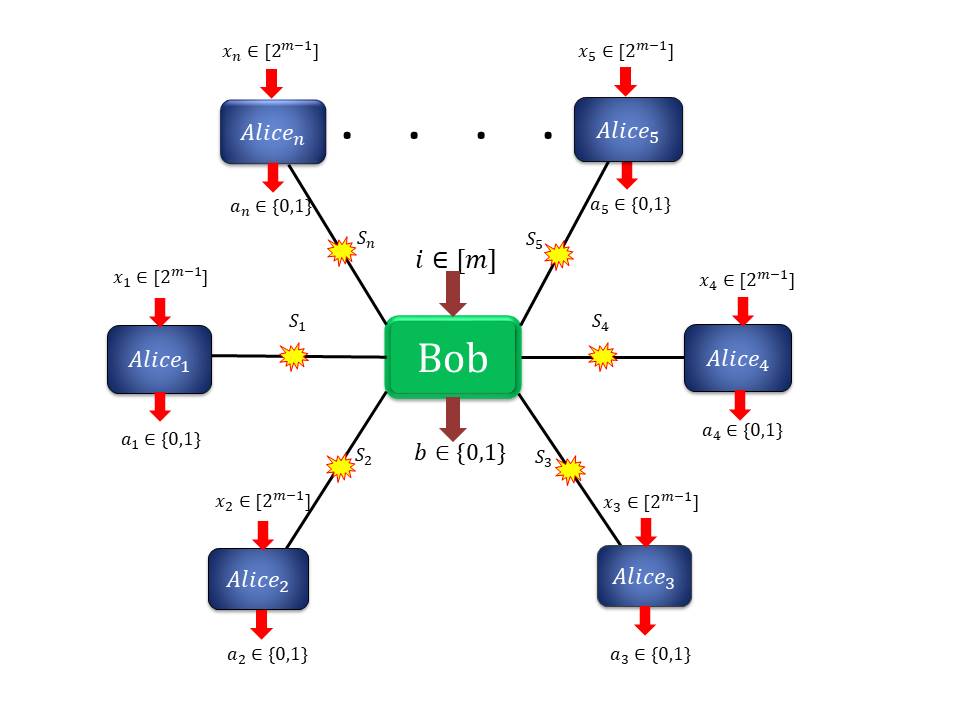}
				\caption{$n$-locality Scenario}\end{center}
			\end{figure} 
Let us first consider the star-network configuration \cite{Armi2014} for $m=3$ which features  total $(n+1)$ parties including $n $ number of edge party (Alices), say Alice$_k,$  $k\in[n]$, the central party Bob and $n$ independent sources  $S_{k}$. For $m=3$, each Alice$_k$ performs the measurements of four binary outcome measurements $A^{k}_{x_{k}}$ according to the  inputs  $x_{k}\in[4]$  and gets outputs $a_{k}\in \{0,1\}$. On the other hand,  Bob performs three binary outcome measurements on the joint system he receives from $n$ sources, and obtains output $b\in \{0,1\}$.  Since, the sources $S_{k}$s are assumed to be independent, Eq. (\ref{fac}) constitutes the $n$-locality assumption. Here  the joint probability distribution in Eq. (\ref{facn}) holds.
Now let us assume that the joint  probabilities are constrained by the following condition:
\ba\label{cond3prob} &&P(a_k=0,b,\alpha_k|x_k=1,i,\chi_k)+\sum\limits_{x_k=2,3,4}P(a_k=1,b,\alpha_k|x_k,i,\chi_k)\hspace{5mm}\\\na
&&=P(a_k=1,b,\alpha_k|x_k=1,i,\chi_k)+\sum\limits_{x_k=2,3,4}P(a_k=0,b,\alpha_k|x_k,i,\chi_k)\ea
where for notational convenience, we denote that the set $\alpha_k$ as the following collection of elements   $\alpha_k=\{a_1,a_2,\dots, a_{k-1},a_{k+1},\dots, a_n\}$ and $\chi_k=\{x_1,x_2,\dots, x_{k-1},x_{k+1},\dots, x_n\}$ and it holds $\forall i\in[3], k\in[n]$.
From Eq. (\ref{cond3prob}), we get that for each edge party  Alice$_k$,   the observables are  constrained by  the following condition:
\ba\label{cond3}A^{k}_{1}-A^{k}_{2}-A^{k}_{3}-A^{k}_{4}=0, \quad\forall k\in[n]\ea
We propose that the following inequality 
	\ba  \label{deltan3}
	(\Delta_{3}^{n})_{nl}=|{I}^{n}_{3,1}|^{\frac{1}{n}}+|{I}^{n}_{3,2}|^{\frac{1}{n}}+|{I}^{n}_{3,3}|^{\frac{1}{n}}\leq 4
	\ea
	is satisfied. Here  '$nl$ ' denotes $n$-locality. Here, $I^{n}_{3,1}$ and $I^{n}_{3,2}$, $I^{n}_{3,3}$  are the linear combinations of suitably chosen correlations defined as 
	\begin{eqnarray}
	\label{pncn31}\nonumber
	I^{n}_{3,1}&=&\bigg\langle\prod\limits_{k=1}^{n}(A^{k}_{1}+A^{k}_{2}+A^{k}_{3}-A^{k}_{4}) B_{1}\bigg\rangle
	\\ \hspace{2mm}
I^{n}_{3,2}&=&\bigg\langle\prod\limits_{k=1}^{n}(A^{k}_{1}+A^{k}_{2}-A^{k}_{3}+A^{k}_{4}) B_{2}\bigg\rangle\\\nonumber
	I^{n}_{3,3}&=&\bigg\langle\prod\limits_{k=1}^{n}(A^{k}_{1}-A^{k}_{2}+A^{k}_{3}+A^{k}_{4}) B_{3}\bigg\rangle
	\end{eqnarray}
	where $A^{k}_{x_{k}}$ denotes observables corresponding to the input $x_{k}\in [4]$ of the $k^{th}$ Alice and 
	\ba
	\nonumber
	\langle{A^{1}_{x_1}\cdots A^{n}_{x_n}B_{i}}\rangle = \sum\limits_{a_{1},..., a_{n},b}(-1)^{\sum_{k=1}^{n}a_{k}+b}P(a_{1},\cdots, a_{n},b|x_{1},\cdots., x_{n},i)\\
	\ea 
	By defining $\langle{A^{k}_{x_{k}}}\rangle_{\lambda_{k}} = \sum\limits_{a_{k}}(-1)^{a_{k}}  P(a_{k}|x_{k},\lambda_{k})$ where $k\in[n]$ and $x_{k}\in [4]\}$ and using the fact that  $|\langle{B_{1}}\rangle_{\lambda_{1},\cdots \lambda_{k}}|\leq{1},$ and the sources are   independent, we can write
	\begin{eqnarray}
	\label{facn31}
	|I^{n}_{3,1}|&\leq&\prod\limits_{k=1}^{n}\bigg|\langle A^{k}_{1}\rangle_{\lambda_k}+\langle A^{k}_{2}\rangle_{\lambda_k}+\langle A^{k}_{3}\rangle_{\lambda_k}-\langle A^{k}_{4}\rangle_{\lambda_k}\bigg|
	\end{eqnarray}
	Similarly, we can  factorize $|I^{n}_{3,2}|$, $|I^{n}_{3,3}|$ and $|I^{n}_{3,4}|$ as Eq. (\ref{facn31}). For our purpose let  $z^1_k=|A^{k}_{1}+A^{k}_{2}+A^{k}_{3}-A^{k}_{4}|,z^2_k=|A^{k}_{1}+A^{k}_{2}-A^{k}_{3}+A^{k}_{4}|$, and  $ z^3_k=|A^{k}_{1}-A^{k}_{2}+A^{k}_{3}+A^{k}_{4}|$.   Now, using the inequality \begin{equation}
	\label{Tavakoli}
	\ \forall\ \ z_{k}^{i} \geq 0; \ \ \ \sum\limits_{i=1}^{m}\bigg(\prod\limits_{k=1}^{n}z_{k}^{i}\bigg)^{\frac{1}{n}}\leq \prod \limits_{k=1}^{n}\bigg(\sum\limits_{i=1}^{m}z_{k}^{i}\bigg)^{\frac{1}{n}}
	\end{equation} for $m=3$, we have
	\ba\bigg(\prod\limits_{k=1}^{n}z_{k}^{1}\bigg)^{\frac{1}{n}}+\bigg(\prod\limits_{k=1}^{n}z_{k}^{2}\bigg)^{\frac{1}{n}}+\bigg(\prod\limits_{k=1}^{n}z_{k}^{3}\bigg)^{\frac{1}{n}}\leq\prod\limits_{k=1}^{n} \bigg(z_{k}^{1}+z_k^2+z^3_k\bigg)^{\frac{1}{n}}\ea
	Substituting the values of $z_k^i$, we get 
	\begin{eqnarray}\label{m=3}
	\nonumber
	\bigg(\prod\limits_{k=1}^{n}|A^{k}_{1}+A^{k}_{2}+A^{k}_{3}-A^{k}_{4}|\bigg)^{\frac{1}{n}}&+&\bigg(\prod\limits_{k=1}^{n}|A^{k}_{1}+A^{k}_{2}-A^{k}_{3}+A^{k}_{4}|\bigg)^{\frac{1}{n}}\\\nonumber
	+\bigg(\prod\limits_{k=1}^{n}|A^{k}_{1}-A^{k}_{2}+A^{k}_{3}+A^{k}_{4}|\bigg)^{\frac{1}{n}}&\leq& \prod\limits_{k=1}^{n}\bigg(|A^{k}_{1}+A^{k}_{2}+A^{k}_{3}-A^{k}_{4}|\\
	+|A^{k}_{1}+A^{k}_{2}-A^{k}_{3}+A^{k}_{4}|&+&|A^{k}_{1}-A^{k}_{2}+A^{k}_{3}+A^{k}_{4}|\bigg)^{\frac{1}{n}}\hspace{9pt}
		\end{eqnarray}
Let us denote   \begin{equation}|A^{k}_{1}+A^{k}_{2}+A^{k}_{3}-A^{k}_{4}|+|A^{k}_{1}+A^{k}_{2}-A^{k}_{3}+A^{k}_{4}|+|A^{k}_{1}-A^{k}_{2}+A^{k}_{3}+A^{k}_{4}|=\eta^k_3\end{equation}
Hence Eq.(\ref{m=3}) provides
	     \begin{eqnarray}
		\label{deltan3pnc}
		(\Delta^{n}_{3})_{nl}&\leq&\prod\limits_{k=1}^{n}\big(\eta^k_3\big)^{\frac{1}{n}}\hspace{1pt}
		\end{eqnarray}
		
	Since the observables $A^k_{x_k}$ are dichotomic and they are bounded by the constraint  	of  Eq. (\ref{cond3}), we get \begin{equation}\bigg[|A^{k}_{1}+A^{k}_{2}+A^{k}_{3}-A^{k}_{4}|+|A^{k}_{1}+A^{k}_{2}-A^{k}_{3}+A^{k}_{4}|+|A^{k}_{1}-A^{k}_{2}+A^{k}_{3}+A^{k}_{4}|\bigg]\leq 4\end{equation}
  Substituting in Eq.(\ref{deltan3pnc}), we finally obtain $(\Delta^{n}_{3})_{nl}\leq 4$ as claimed in Eq. (\ref{deltan3}).
		
	
	To derive the quantum value of $(\Delta^{n}_{3})_{Q}$ we consider that the sources emit independent quantum systems. We introduce an elegant SOS approach to derive  the optimal quantum value of  $(\Delta^{n}_{3})_{Q}$. We essentially show that there is a positive semidefinite operator $\langle\gamma^{n}_{3}\rangle\geq 0$ that can be expressed as  
	\ba  
	\langle \gamma^n_{3}\rangle_{Q}=-(\Delta^{n}_{3})_{Q}+\beta^{n}_{3}
	\ea 
	where $\beta^{n}_{3}$ is the optimal value which can be obtained when $\langle \gamma^n_{3}\rangle_{Q}$ is equal to zero. To prove this, let us  consider a set of suitable positive numbers $|M^{n}_{3,i}|\psi\rangle|$ which is polynomial functions of $A^{k}_{x_{k}}, (\forall k)$ and $B_{i}$  so that 
	\begin{eqnarray}
	\label{gamma3}\na
	\langle\gamma^{n}_{3}\rangle&=&\dfrac{(\omega_{3,1}^{n})^{\frac{1}{n}}}{2}|M^{n}_{3,1}|\psi\rangle|^2+ \frac{(\omega_{3,2}^{n})^{\frac{1}{n}}}{2}|M^{n}_{3,2}|\psi\rangle|^2+\dfrac{(\omega_{3,3}^{n})^{\frac{1}{n}}}{2}|M^{n}_{3,3}|\psi\rangle|^2\\\end{eqnarray}
	where $\omega^{n}_{3,i}$ is defined as suitable positive numbers such that $\omega^{n}_{3,i}=\prod\limits_{k=1}^{n}(\omega^{n}_{3,i})_{A_{k}}$. Clearly, $(\Delta^{n}_{3})^{opt}_{Q}$ is obtained if $\langle \gamma^{n}_{3}\rangle_{Q}=0$ i.e.,  	$|M^{n}_{3,i}|\psi\rangle|=0$, $\forall i\in\{1,2,3\}$ .
	For notational convenience,  we denote $|\psi\rangle_{A_{1}A_{2}\cdots A_nB}=|\psi\rangle$.

	 If we consider the term, $\bigg|\prod\limits_{k=1}^{n}\left(\frac{{A}^{k}_{1}+{A}^{k}_{2}+A^{k}_{3}-{A}^{k}_{4}}{(\omega_{3,1}^{n})_{A_{k}}}\right) |\psi\rangle
	\bigg|^{\frac{1}{n}} -|B_{1}|\psi\rangle|^{\frac{1}{n}}$; clearly, the operator part of the first  term is normalized, which implies that the first term gives    $1$. Since $B_1$ is a dichotomic operator, $|B_{1}|\psi\rangle|^{\frac{1}{n}}$ must be less than or equals to $1$ which implies that the complete term is positive. Similar logic holds for the other terms as well. Hence we can define   the positive number $|M^{n}_{3,1}|\psi\rangle|$, $|M^{n}_{3,2}|\psi\rangle|$  and $|M^{n}_{3,3}|\psi\rangle|$   as follows:

\ba\nonumber
\label{m1m2m3}
    |M^{n}_{3,1}|\psi\rangle|&=&\bigg|\prod\limits_{k=1}^{n}\left(\frac{{A}^{k}_{1}+{A}^{k}_{2}+A^{k}_{3}-A^{k}_{4}}{(\omega_{3,1}^{n})_{A_{k}}}\right) |\psi\rangle
	\bigg|^{\frac{1}{n}} -|B_{1}|\psi\rangle|^{\frac{1}{n}}\\ 
	|M^{n}_{3,2}|\psi\rangle|&=&\bigg|\prod\limits_{k=1}^{n}\left(\frac{{A}^{k}_{1}+{A}^{k}_{2}-A^{k}_{3}+A^{k}_{4}}{(\omega_{3,2}^{n})_{A_{k}}}\right) |\psi\rangle	\bigg|^{\frac{1}{n}} -|B_{2}|\psi\rangle|^{\frac{1}{n}}\hspace{3pt}\\\nonumber	|M^{n}_{3,3}|\psi\rangle|&=&\bigg|\prod\limits_{k=1}^{n}\left(\frac{{A}^{k}_{1}-{A}^{k}_{2}+A^{k}_{3}+A^{k}_{4}}{(\omega_{3,3}^{n})_{A_{k}}}\right) |\psi\rangle
	\bigg|^{\frac{1}{n}} -|B_{3}|\psi\rangle|^{\frac{1}{n}}\ea
Substituting these in Eq. (\ref{gamma3}), we get
	\ba\langle\gamma^{n}_{3}\rangle_{Q}=-(\Delta^{n}_{3})_{Q} +\sum\limits_{i=1}^{3}(\omega^{n}_{3,i})^{\frac{1}{n}}\ea 	The optimal value of $(\Delta^{n}_{3})_{Q}$ is obtained if  $\langle \gamma^n_{3}\rangle_{Q}=0$.	Hence,
	\begin{eqnarray}(\Delta^{n}_{3})_{Q}^{opt} =max\left(\sum\limits_{i=1}^{3}(\omega^{n}_{3,i})^{\frac{1}{n}}\right)
	\end{eqnarray}
	where
	\ba
	&&(\omega^{n}_{3,1})_{A_{k}}=||A^{k}_{1}+A^{k}_{2}+A^{k}_{3}-A^{k}_{4}||_{2}\\
	\nonumber
	&=&\Big(4+\langle \{A^{k}_{1},(A^{k}_{2}+A^{k}_{3}-A^{k}_{4})\}+\{A^{k}_{2},(A^{k}_{3}-A^{k}_{4})\}-\{A^{k}_{3},A^{k}_{4}\}\rangle\Big)^{1/2}	\ea
	
	Similarly,  we can write for
	$(\omega^{n}_{3,2})_{A_{k}}$ and $ (\omega^{n}_{3,3})_{A_{k}}$, $\forall k\in[n]$. Since  $\omega^{n}_{3,i}=\prod\limits_{k=1}^{n}(\omega^{n}_{3,i})_{A_{k}}$, by using the  inequality  Eq. (\ref{Tavakoli}),
	we get \ba\label{wn3i}\sum\limits_{i=1}^{3}(\omega^{n}_{3,i})^{\frac{1}{n}}\leq\prod\limits_{k=1}^{n}\bigg( \sum\limits_{i=1}^{3}(\omega^{n}_{3,i})_{A_{k}}\bigg)^{\frac{1}{n}}\ea
	Further using the convex inequality, we have  \ba\label{convex3}\sum\limits_{i=1}^{3}(\omega^{n}_{3,i})_{A_{k}}\leq\sqrt{ 3\sum\limits_{i=1}^{3}\bigg((\omega^{2}_{3,i})_{A_{k}}\bigg)^2}\hspace{3pt}\ea 
 The equality holds when each of $(\omega^{n}_{3,i})_{A_k}$ is equal to each other.	We can then write  	\begin{eqnarray}
		\label{sumw23ik}\nonumber
		\sum\limits_{i=1}^{3}\bigg((\omega^{n}_{3,i})_{A_{k}}\bigg)^2&=& \langle\psi|12+(\{A^{k}_{1},(A^{k}_{2}+A^{k}_{3}+A^{k}_{4})\}-\{A^{k}_{2},(A^{k}_{3}+A^{k}_{4})\}\\&&-\{A^{k}_{3},A^{k}_{4}\})|\psi\rangle=\langle\psi(12+\delta_3)|\psi\rangle 
  \end{eqnarray} where \ba\nonumber\delta_3=(\{A^{k}_{1},(A^{k}_{2}+A^{k}_{3}+A^{k}_{4})\}-\{A^{k}_{2},(A^{k}_{3}+A^{k}_{4})\}-\{A^{k}_{3},A^{k}_{4}\})\\\ea

		Let $|\psi'\rangle=(A^{k}_{1}-A^{k}_{2}-A^{k}_{3}-A^{k}_{4})|\psi\rangle$  such that $|\psi\rangle\neq 0$. Then
		$\langle\psi'|\psi'\rangle=\langle\psi|(4-\delta_3)|\psi\rangle$ implies $\langle\delta_3\rangle=4-\langle\psi'|\psi'\rangle$.
Evidently, $\langle\delta_3\rangle_{max}$ is obtained only when $\langle\psi'|\psi'\rangle=0$. Since, $|\psi\rangle\neq 0$, we then have 

\ba\label{c3}\label{c3}A^{k}_{1}-A^{k}_{2}-A^{k}_{3}-A^{k}_{4}=0\ea

	Hence, to obtain the optimal value of  $(\Delta^{n}_{3})_{Q}^{opt}$, observables of each Alice$_k$ must satisfy  the linear condition of Eq.(\ref{c3}).

	In turn  $\langle\delta_3\rangle_{max}=4$  provides $\sum\limits_{i=1}^{3}\bigg((\omega^{n}_{3,i})_{A_{k}}\bigg)^2=16$. Plugging it in the above mentioned  inequality (\ref{convex3}), we get $\sum\limits_{i=1}^{3}(\omega^{n}_{3,i})_{A_k}\leq 4\sqrt{3}$. Since each observable $A^k_{x_k}$ is dichotomic, pre-multiplying and post-multiplying Eq.(\ref{c3}) by $A^k_1$, and adding them we 

	 \ba \label{AK31}1- A^k_1A^k_2- A^k_1A^k_3- A^k_1A^k_4&=0\\	 \label{AK32}1- A^k_2A^k_1- A^k_3A^k_1- A^k_4A^k_1&=0\ea	 Adding Eq.(\ref{AK31}) and Eq.(\ref{AK32}),	 we get \ba\label{AK3}\{A^k_1,A^k_2\}+\{A^k_1,A^k_3\}+\{A^k_1,A^k_4\}=2\ea
	 Similarly, three more such relations can be found which  are the following:
	 \ba\label{c3d}\nonumber
	 \{A^k_1,A^k_2\}-\{A^k_2,A^k_3\}-\{A^k_2,A^k_4\}&=&2
	\\
	\{A^k_1,A^k_3\}-\{A^k_2,A^k_3\}-\{A^k_3,A^k_4\}&=&2\\\nonumber
	\{A^k_1,A^k_4\}-\{A^k_2,A^k_4\}-\{A^k_3,A^k_4\}&=&2\ea
	Solving  Eqs. (\ref{AK3}-\ref{c3d}), $ \forall k\in[n]$, we get 
 \ba\label{aanti3}
 &&\{A^{k}_{1}, A^{k}_{2}\}=\{A^{k}_{1}, A^{k}_{3}\}=\{A^{k}_{1}, A^{k}_{4}\}=\frac{2}{3}\\
 \nonumber
 &&\{A^{k}_{2}, A^{k}_{3}\}=\{A^{k}_{2}, A^{k}_{4}\}=\{A^{k}_{3}, A^{k}_{4}\}=-\frac{2}{3}.\ea   We thus obtain the relations between the  observables for each Alice$_{k}$. Also, for optimal value, we check that 
 \ba(\omega^{n}_{3,1})_{A_{k}}=(\omega^{n}_{3,2})_{A_{k}}=(\omega^{n}_{3,3})_{A_{k}}=4/\sqrt{3}\ea This, in turn, provides the optimal quantum value
 \ba(\Delta^{n}_{3})_{Q}^{opt}= 4\sqrt{3}\ea
 
  \section{ $n$-locality scenario in star-network for $m=4$} \label{IV} Before presenting the result for arbitrary $m$, let us demonstrate the result in $n$-local scenario for $m=4$  where  Bob performs  four measurements  $B_{i}$ according to the inputs ${i}\in\{1,2,3,4\}$  and gets output $b\in \{0,1\}$. Each of Alice$_{k},$ $ k\in[n]$  performs the measurements of eight dichotomic observables denoted by $A^{k}_{x_{k}}$, according to the inputs $x_{k}\in[8]$  and gets output $a_k\in\{0,1\}$.
  Now let us assume that the joint  probabilities are constrained by the following conditions:\begin{widetext}
\ba\label{cond4prob}\na 
\sum\limits_{x_k=1,2,8}P(a_k=0,b,\alpha_k|x_k,i,\chi_k)+\sum\limits_{x_k=3,4,5,6,7}P(a_k=1,b,\alpha_k|x_k,i,\chi_k)&=&\sum\limits_{x_k=1,2,8}P(a_k=1,b,\alpha_k|x_k,i,\chi_k)+\sum\limits_{x_k=3,4,5,6,7}P(a_k=0,b,\alpha_k|x_k,i,\chi_k),\\\na
\sum\limits_{x_k=1,3,7}P(a_k=0,b,\alpha_k|x_k,i,\chi_k)+\sum\limits_{x_k=2,4,5,6,8}P(a_k=1,b,\alpha_k|x_k,i,\chi_k)&=&\sum\limits_{x_k=1,3,7}P(a_k=1,b,\alpha_k|x_k,i,\chi_k)+\sum\limits_{x_k=2,4,5,6,8}P(a_k=0,b,\alpha_k|x_k,i,\chi_k),\\\na
\sum\limits_{x_k=1,4,6}P(a_k=0,b,\alpha_k|x_k,i,\chi_k)+\sum\limits_{x_k=2,3,5,7,8}P(a_k=1,b,\alpha_k|x_k,i,\chi_k)&=&\sum\limits_{x_k=1,4,6}P(a_k=1,b,\alpha_k|x_k,i,\chi_k)+\sum\limits_{x_k=2,3,5,7,8}P(a_k=0,b,\alpha_k|x_k,i,\chi_k),\\\na
\sum\limits_{x_k=1,5,6,7,8}P(a_k=0,b,\alpha_k|x_k,i,\chi_k)+\sum\limits_{x_k=2,3,4}P(a_k=1,b,\alpha_k|x_k,i,\chi_k)&=&\sum\limits_{x_k=1,5,6,7,8}P(a_k=1,b,\alpha_k|x_k,i,\chi_k)+\sum\limits_{x_k=2,3,4}P(a_k=0,b,\alpha_k|x_k,i,\chi_k)\\
\ea
\end{widetext}
where for notational convenience, we denote that the set $\alpha_k$ as the following collection of elements   $\alpha_k=\{a_1,a_2,\dots, a_{k-1},a_{k+1},\dots, a_n\}$ and $\chi_k=\{x_1,x_2,\dots, x_{k-1},x_{k+1},\dots, x_n\}$ and it holds $\forall i\in[4], k\in[n]$.
From Eq. (\ref{cond4prob}), we get that for each edge party  Alice$_k$,   the observables are  constrained by  the following conditions:
\begin{eqnarray}\label{cond4}\na
		A^{k}_{1}+A^{k}_{2}-A^{k
		}_{3}-A^{k}_{4}-A^{k}_{5}-A^{k
		}_{6}-A^{k}_{7}+A^{k}_{8}&=&0\\
		\nonumber
		A^{k}_{1}-A^{k}_{2}+A^{k
		}_{3}-A^{k}_{4}-A^{k}_{5}-A^{k
		}_{6}+A^{k}_{7}-A^{k}_{8}&=&0\\
		\nonumber
		A^{k}_{1}-A^{k}_{2}-A^{k
		}_{3}+A^{k}_{4}-A^{k}_{5}+A^{k
		}_{6}-A^{k}_{7}-A^{k}_{8}&=&0\\
		A^{k}_{1}-A^{k}_{2}-A^{k
		}_{3}-A^{k}_{4}+A^{k}_{5}+A^{k
		}_{6}+A^{k}_{7}+A^{k}_{8}&=&0
		\end{eqnarray}
 We propose the following inequality is given by  
		\begin{equation}
		\label{deltapncn4}
		(\Delta_{4}^{n})_{nl} = |{I}^{n}_{4,1}|^{\frac{1}{n}}+|{I}^{n}_{4,2}|^{\frac{1}{n}}+|{I}^{n}_{4,3}|^{\frac{1}{n}}+|{I}^{n}_{4,4}|^{\frac{1}{n}}\leq 8
		\end{equation} where we define $I_{4,1}^{n},I_{4,2}^{n},I_{4,3}^{n},I_{4,4}^{n} $ are suitable linear combinations of correlations defined as follows:		\begin{eqnarray}\nonumber
I_{4,1}^{n}&=&\bigg\langle\prod\limits_{k=1}^{n}(A^{k}_{1}+A^{k}_{2}+A^{k
		}_{3}+A^{k}_{4}-A^{k}_{5}+A^{k
		}_{6}+A^{k}_{7}+A^{k}_{8})B_{1}\bigg\rangle\\\na
	 I_{4,2}^{n}&=&\bigg\langle\prod\limits_{k=1}^{n}(A^{k}_{1}+A^{k}_{2}+A^{k
		}_{3}-A^{k}_{4}+A^{k}_{5}+A^{k
		}_{6}-A^{k}_{7}-A^{k}_{8})B_{2}\bigg\rangle\hspace{10pt}\\\nonumber 
		I_{4,3}^{n}&=&\bigg\langle\prod\limits_{k=1}^{n}(A^{k}_{1}+A^{k}_{2}-A^{k
		}_{3}+A^{k}_{4}+A^{k}_{5}-A^{k
		}_{6}+A^{k}_{7}-A^{k}_{8})B_{3}\bigg\rangle\\\nonumber I_{4,4}^{n}&=&\bigg\langle\prod\limits_{k=1}^{n}(A^{k}_{1}-A^{k}_{2}+A^{k
		}_{3}+A^{k}_{4}+A^{k}_{5}-A^{k
		}_{6}-A^{k}_{7}+A^{k}_{8})B_{4}\bigg\rangle\\
		\end{eqnarray}
		Again using $n$-locality assumption, we can factorize $|I^{n}_{4,1}|$, $|I^{n}_{4,2}|$ and $|I^{n}_{4,3}|$, $|I^{n}_{4,4}|$  as in Eq. (\ref{facn31}). By using the inequality  (\ref{Tavakoli}) for $m=4$ and by following a similar method as stated above, we can write 
	\begin{equation}\label{dn4}
(\Delta_{4}^{n})_{nl}\leq \prod\limits_{k=1}^{n}(\eta_4^k)^{\frac{1}{n}}
	\end{equation}
	 where $\eta_4^k=|A^{k}_{1}+A^{k}_{2}+A^{k
		}_{3}+A^{k}_{4}-A^{k}_{5}+A^{k
		}_{6}+A^{k}_{7}+A^{k}_{8}|+|A^{k}_{1}+A^{k}_{2}+A^{k
		}_{3}-A^{k}_{4}+A^{k}_{5}+A^{k
		}_{6}-A^{k}_{7}-A^{k}_{8}|+|A^{k}_{1}+A^{k}_{2}-A^{k
		}_{3}+A^{k}_{4}+A^{k}_{5}-A^{k
		}_{6}+A^{k}_{7}-A^{k}_{8}|+|A^{k}_{1}-A^{k}_{2}+A^{k
		}_{3}+A^{k}_{4}+A^{k}_{5}-A^{k
		}_{6}-A^{k}_{7}+A^{k}_{8}|$. 
   
   Since, each observable $A^k_{x_k}$ is dichotomic and bounded by the  constraints Eq. (\ref{cond4}), we get 
   that $\eta_4^k\leq 8, \forall k\in[n]$. Substituting it in Eq. (\ref{dn4}), clearly we get 
   the inequality in  Eq. (\ref{deltapncn4}). The optimal quantum value of $(\Delta_{4}^{n})_{Q}^{opt}=16>(\Delta_{4}^{n})_{nl}$. We derived the nature of state and observables required for this optimization which is again independent of the dimension of the system. The detailed derivation is quite lengthy and thus deferred to  Appendix \ref{A}.

   We have also provided the detailed derivation of $n$-locality inequality and its optimal quantum violation for $m=5$ in Appendix \ref{B}. 
	 	
\section{ $n$-locality scenario in star-network for arbitrary $m$ inputs}\label{V}  We  generalize the $n$-locality inequality  for arbitrary $m$ number of inputs for  the central party Bob.  Here each of Alice$_{k} ( k\in[n]$)  performs the measurements of $2^{m-1}$ dichotomic observables denoted by $A^{k}_{x_{k}}$, according to the inputs $x_{k}\in[2^{m-1}]$  and obtains output  $a_{k}\in \{0,1\}$.

We propose the following $n$-locality inequality   
		\begin{equation}
		\label{deltapncnm}
		(\Delta_{m}^{n})_{nl} =\sum\limits_{i=1}^{m} |{I}^{n}_{m,i}|^{\frac{1}{n}} \leq 2^{m-1}
		\end{equation} where we define $I_{m,i}^{n}$ as suitable linear combination of correlaltions as
		\begin{eqnarray}\label{Inmi}
{I}^{n}_{m,i}=\bigg\langle\prod\limits_{k=1}^{n}\sum\limits_{x_{k}=1}^{2^{m-1}}(-1)^{y^{x_{k}}_{i}} A_{x_{k}}^{k}B_{i}\bigg\rangle
		\end{eqnarray}
		Here ${y^{x_{k}}_{i}}$ takes value either $0$ or $1$, for each $k\in[n]$.  For our purpose, we fix the values of ${y^{x_{k}}_{i}}$ by using  the encoding scheme used in Random Access Codes (RACs) \cite{Ambainis,Ghorai2018,AKP2020,Asmita} as a tool. This will fix $1$ or $-1$ values of $(-1)^{y^{x_{k}}_{i}}$ in Eq. (\ref{Inmi}) for a given $i$. Let us consider a random variable $y^{\alpha}\in \{0,1\}^{m}$ with $\alpha\in \{1,2...2^{m}\}$. Each element of the bit string can be written as $y^{\alpha}=y^{\alpha}_{i=1} y^{\alpha}_{i=2} y^{\alpha}_{i=3} .... y^{\alpha}_{i=m}$. For  example, if $y^{\alpha} = 011...00$ then $y^{{\alpha}}_{i=1} =0$, $y^{{\alpha}}_{i=2} =1$, $y^{{\alpha}}_{i=3} =1$ and so on. Here we  denote  the length $m$ binary strings as $y^{x_k}$.   Here we consider the bit strings such that for any two $k$ and $k'$,  $y^{x_k}\oplus_2y^{x_k'}=11\cdots1$. Clearly, we have $x_k\in \{1,2...2^{m-1}\}$ constituting the inputs for Alice $_k$. If $x_k=1$, we get all the first bit of each  bit  string $y_i$ for every $i\in \{1,2 \cdots m\}$.

	Since $|B_i|\leq 1, \forall i\in[m]$, from Eq. (\ref{Inmi}) we can write
		\begin{eqnarray}\label{Inmi2}
|{I}^{n}_{m,i}|\leq \prod\limits_{k=1}^{n}\bigg|\bigg\langle\sum\limits_{x_{k}=1}^{2^{m-1}}(-1)^{y^{x_{k}}_{i}} A_{x_{k}}^{k}\bigg\rangle\bigg|
		\end{eqnarray}
By using the inequality (\ref{Tavakoli}),  we get  
		\begin{eqnarray}\label{deltanmleq}
	(\Delta_{m}^{n})_{nl}    \leq\prod\limits_{k=1}^{n}\bigg(\sum\limits_{i=1}^{m} \bigg|\bigg\langle\sum\limits_{x_{k}=1}^{2^{m-1}}(-1)^{y^{x_{k}}_{i}} A_{x_{k}}^{k}\bigg\rangle\bigg|\bigg)^{\frac{1}{n}}\leq\prod\limits_{k=1}^{n}(\eta^k_m)^{\frac{1}{n}}
		\end{eqnarray}  
		where we  denote  \ba\eta^k_m=\sum\limits_{i=1}^{m}\bigg|\bigg\langle\sum\limits_{x_{k}=1}^{2^{m-1}}(-1)^{y^{x_{k}}_{i}} A_{x_{k}}^{k}\bigg\rangle\bigg|\ea
Now we  consider a set $\mathcal{L}_m=\{s|s\in\{0,1\}^m, \sum_r s_r\geq 2\}$, $r\in\{1,2,\cdots m\}$. The element  $s_l\in \mathcal{L}_m$ is  such that $(s_l)_r\neq 2u$, for some $u\in\mathbb{N}$. We then find $(2^{m-1}-m)$ number of such elements  $s_l$  where $l\in[2^{m-1}-m]$. 		
Let us assume that the joint  probabilities are constrained by the following conditions: 
\ba\label{conditionmprob}\na
\sum\limits_{x_k\in U_{k,l}} P(a_k=0,b,\alpha_k|x_k,i,\chi_k)+\sum\limits_{x_k\notin U_{k,l}} P(a_k=1,b,\alpha_k|x_k,i,\chi_k)=\\\na
\sum\limits_{x_k\in U_{k,l}} P(a_k=1,b,\alpha_k|x_k,i,\chi_k)+\sum\limits_{x_k\notin U_{k,l}} P(a_k=0,b,\alpha_k|x_k,i,\chi_k)\\\ea
 where we define the set $U_{k,l}\subset [2^{m-1}]$ as the collection of $x_k$s  such that for a given $s_l$,  $(-1)^{s_l.y^{x_k}}=1$ i.e., 
 \ba U_{k,l}=\{x_k:(-1)^{s_l.y^{x_k}}=1\}\ea Also for notational convenience, we denote that the set $\alpha_k$ as the following collection of elements   $\alpha_k=\{a_1,a_2,\dots, a_{k-1},a_{k+1},\dots, a_n\}$ and $\chi_k=\{x_1,x_2,\dots, x_{k-1},x_{k+1},\dots, x_n\}$ and it holds $\forall i\in[m], k\in[n]$.
Since there are total  $(2^{m-1}-m)$ number of elements $s_l$, from Eq. (\ref{conditionmprob}), we get that the observables  for each edge party  Alice$_k$  constrained by  the following $(2^{m-1}-m)$ conditions:
\begin{eqnarray}\label{conditionm}
\sum\limits_{x_{k}=1}^{2^{m-1}} (-1)^{s_l.y^{x_{k}}}A^k_{x_k}=0, \forall s_l\in[2^{m-1}-m],k\in[n].\ea

 Since, each observable $A^k_{x_k}$ is dichotomic and bounded by the  constraints Eq. (\ref{conditionm}),  these  in turn provides the inequality in  Eq. (\ref{deltapncnm}) \cite{Ghorai2018,AKP2020}.

To find the optimal quantum value of the expression $(\Delta_{m}^{n})$,  we use an elegant form of SOS approach,  so that, $(\Delta_{m}^{n})_{Q}\leq \beta_{m}^{n}$  for all possible quantum states $\rho_{A_{k}B}$  and measurement operators $A^{k}_{x_{k}}$ and $B_{i}$. Here $\beta_{m}^{n}$ is the optimal quantum value of $(\Delta_{m}^{n})_{Q}$. This is equivalent of showing that there is a positive semidefinite operator $\langle\gamma^{n}_{m}\rangle\geq 0$, which can be expressed as \ba \langle \gamma_{m}^{n}\rangle_{Q}=-(\Delta^{n}_{m})_{Q}+\beta^{n}_{m}.\ea This can be proven by considering a set of suitable positive operators $M^{n}_{m,i}$ which is polynomial functions of   $A^{k}_{x_{k}}$ and $B_{i}$,   given by
	\ba
	\label{gammanm}
	\langle\gamma^{n}_{m}\rangle=\sum\limits_{i=1}^{m}  \frac{{(\omega^{n}_{m,i}})^{\frac{1}{n}}}{2 }|M^{n}_{m,i}|\psi\rangle|^2
	\ea
	where $\omega^{n}_{m,i}$ is positive number with $\omega^{n}_{m,i}=\prod \limits _{k=1}^{n}(\omega^{n}_{m,i})_{A_{k}}$. The optimal quantum value of $(\Delta^{n}_{m})_{Q}$ is obtained if $\langle \gamma_{m}^{n}\rangle_{Q}=0$, implying that 
	\begin{align}
	\label{mnmi}
	\forall i, \ \ \ |M^{n}_{m,i}|\psi\rangle_{A_{1}A_{2}\cdots A_{n}B}|=0
	\end{align}
where $|\psi\rangle_{A_{1}A_{2}\cdots A_{n} B}=|\psi\rangle_{A_{1}B}\otimes |\psi\rangle_{A_{2}B}\otimes . . . . \otimes |\psi\rangle_{\cdots A_{n}B}$ and $|\psi\rangle_{A_{k}B}$s are the two-party quantum states originating from independent sources $S_{k}$.
 If we consider the term, $\prod\limits_{k=1}^{n}\frac{1}{(\omega_{m,i}^{n})_{A_{k}}^\frac{1}{n}}\bigg|\bigg[\sum\limits_{i=1}^{m}\sum\limits_{x_{k}=1}^{2^{m-1}}(-1)^{y^{x_{k}}_{i}} A_{x_{k}}^{k}\hspace{1pt}\bigg]|\psi\rangle\bigg|^{\frac{1}{n}} -|B_{i}|\psi\rangle|^{\frac{1}{n}}$; clearly, the operator part of the first  term is normalized, which implies that the first term gives    $1$. Since $B_i$ is a dichotomic operator, $|B_{i}|\psi\rangle|^{\frac{1}{n}}$ must be less than or equals to $1$ which implies that the complete term is positive. To satisfy  the form of Eq. (\ref{gammanm}),   we can suitably  define   the positive number $|M^{n}_{m,i}|\psi\rangle|$ as
	\begin{equation}
	\label{nmi}
	|M^{n}_{m,i}|\psi\rangle|=\prod\limits_{k=1}^{n}\frac{1}{(\omega_{m,i}^{n})_{A_{k}}^\frac{1}{n}}\bigg|\bigg[\sum\limits_{i=1}^{m}\sum\limits_{x_{k}=1}^{2^{m-1}}(-1)^{y^{x_{k}}_{i}} A_{x_{k}}^{k}\hspace{1pt}\bigg]|\psi\rangle\bigg|^{\frac{1}{n}}-|B_{i}|\psi\rangle|^{\frac{1}{n}}
	\end{equation}
	with 
\ba
 \nonumber
(\omega^{n}_{m,i})_{A_{k}}&=&||\langle\psi|\sum\limits_{i=1}^{m}\sum\limits_{x_{k}=1}^{2^{m}-1}(-1)^{y^{x_{k}}_{i}} A_{x_{k}}^{k}\hspace{1pt}|\psi\rangle||_{2}\\&=&\bigg(\langle\psi|\bigg[\sum\limits_{i=1}^{m}\sum\limits_{x_{k}=1}^{2^{m-1}}(-1)^{y^{x_{k}}_{i}} A_{x_{k}}^{k}\hspace{1pt}\bigg]^{\dagger}\bigg[\sum\limits_{i=1}^{m}\sum\limits_{x_{k}=1}^{2^{m-1}}(-1)^{y^{x_{k}}_{i}} A_{x_{k}}^{k}\hspace{1pt}\bigg]|\psi\rangle\bigg)^{\frac{1}{2}}\hspace{15pt}
\ea
Here for simplicity we write $|\psi\rangle_{A_{1}A_{2}\cdots A_{n} B}=|\psi\rangle$. Plugging Eq. (\ref{nmi}) into Eq. (\ref{gammanm}) and by noting that $(A^{k}_{x_{k}})^{\dagger}A^{k}_{x_{k}}=B_{i}^{\dagger} B_{i}=\mathbb{I}$, we get
	\begin{align}
	\langle \gamma^{n}_{m}\rangle_{Q}=-(\Delta^{n}_{m})_{Q} + \sum\limits_{i=1}^{m}(\omega^{n}_{m,i})^{\frac{1}{n}}
	\end{align}
which in turn provides
	
	\begin{align}
	\label{optbnm}
	(\Delta^{n}_{m})_{Q}^{opt} &= \sum\limits_{i=1}^{m}(\omega^{n}_{m,i})^{\frac{1}{n}}
	\end{align} 
	 Using the inequality  (\ref{Tavakoli}),  we can get
	\begin{align} 
	\label{optbnm}
	(\Delta^{n}_{m})_{Q}^{opt} &\leq \prod \limits_{k=1}^{n} \bigg(\sum\limits_{i=1}^{m}(\omega^{n}_{m,i})_{A_k}\bigg)^{\frac{1}{n}}\end{align} 
Using convex inequality, we can write 
\ba\sum\limits_{i=1}^{m}(\omega^{n}_{m,i})_{A_k}\leq\sqrt{m\sum\limits_{i=1}^{m}\bigg((\omega^{n}_{m,i})_{A_k}\bigg)^2}\ea
Here, using the definition of $(\omega^{n}_{m,i})_{A_k}$, we can write   $\sum\limits_{i=1}^{m}\bigg((\omega^{n}_{m,i})_{A_k}\bigg)^2=\langle\psi|(m2^{m-1}+\delta_m)\mathbb{I}|\psi\rangle$ where  \ba\delta_m&=&\sum\limits_{l=1}^{2^{m-1}-m}(\delta_m)_l\\\nonumber
&=&(m-2)\sum\limits_{j'=2}^{1+\binom{m}{1}}\{A^k_1,A^k_{j'}\}+(m-4)\sum\limits_{j'=2+\binom{m}{1}}^{1+\binom{m}{1}+\binom{m}{2}}\{A^k_1,A^k_{j'}\}+\cdots\\\nonumber&&+\bigg(m-2\lfloor\frac{m}{2}\rfloor\bigg)\sum\limits_{j'=2+\binom{m}{1}+\binom{m}{1}+\binom{m}{2}+\cdots\binom{m}{\lfloor\frac{m}{2}\rfloor-1} }^{1+\binom{m}{1}+\binom{m}{1}+\binom{m}{2}+\cdots\binom{m}{\lfloor\frac{m}{2}\rfloor}}\{A^k_1,A^k_{j'}\}+(m-4)\\&&\sum\limits_{j,j'=2, j\neq j'}^{1+\binom{m}{1}}\{A^k_j,A^k_{j'}\}
+\cdots\cdots +(m-4)\{A^k_{2^{m-1}-1},A^k_{2^{m-1}} \}\hspace{18pt}\ea 
such that $(\delta_m)_l=2^{m-1}-\langle\psi_l|\psi_l\rangle$.
Hence, 
\ba \delta_m=(2^{m-1}-m)2^{m-1}-\sum\limits_{l=1}^{2^{m-1}-m}\langle\psi_l|\psi_l\rangle\ea
 where we define 
 
 \ba\label{ll}|\psi_l\rangle=\sum\limits_{x_{k}=1}^{2^{m-1}} (-1)^{s_l.y^{x_{k}}}A^{k}_{x_{k}}|\psi\rangle.\ea 
 
 The element $s_l\in \mathcal{L}_m$ is the same  as defined earlier.  Clearly, 	$(\delta_m)_{max}=(2^{m-1}-m)2^{m-1}$ and it holds only when for each $l\in[2^{m-1}-m]$, $|\psi_l\rangle=0$. Since $|\psi\rangle\neq 0$ , hence for optimization, the observables for each Alice must satisfy the conditions  $\sum\limits_{x_{k}=1}^{2^{m-1}} (-1)^{s_l.y^{x_{k}}}A^{k}_{x_{k}}=0$, for each $l\in[2^{m-1}-m]$.
  Finally, we get  $\sum\limits_{i=1}^{m}\bigg((\omega^{n}_{m,i})_{A_k}\bigg)_{opt}^2=2^{2(m-1)}$ which in turn gives  \ba(\Delta^{n}_{m})_{Q}^{opt}=2^{m-1}\sqrt{m}\ea

	To obtain the optimal quantum value, the observables of each Alice$_k$ need to satisfy the condition
	\ba\label{ajj'}\{A^k_j,A^k_{j'}\}=2-\frac{4p}{m}\ea    where $j,j'=x_k\in[2^{m-1}]$. Clearly, there exists the $j(j')^{th}$ bit string denoted by $y^{j}(y^{{j'}})$ from the set of $2^{m-1}$ bit strings as defined earlier. Let the set  $\{y^{j}\}$ contains all the elements ($0$ or $1$) of that corresponding bit string. 
	Hence, for $x_k=j(j')\in[2^{m-1}]$, we can consider the set $\{y^{j}\}\cup \{y^{{j'}}\}$ as the collection of those elements corresponds to the bit strings   $\{y^{j}\}$ and $\{y^{j'}\}$. Without loss of generality, let us assume, $\{y^{j}\}\cup \{y^{{j'}}\}$ contains  $q$ number of $1$s in it. Clearly, from the construction of the bit strings, here $0\leq q\leq m$.  
	
	We can divide  the bit strings $y^{j}(y^{j'})$ into $\lfloor\frac{m}{2}\rfloor$ classes according to the number of $1$s in it. Let $y^{j}\in C^{\nu}$ if the corresponding bit string of $y^{j}$ contains $\nu$ number of $1$s in it.  Let there are two classes $C^{\nu}$ and $C^{\nu'}$ such that $y^{j}\in C^{\nu}$ and $y^{j'}\in C^{\nu'}$ and $\nu+\nu'=q$ ($0\leq \nu, \nu'\leq\lfloor\frac{m}{2}\rfloor$). For a given pair  $(j,j')$, there exists $t\in \mathbb{T}\subseteq [m] $ such that $y^j_t=y^{j'}_t=1$. Let the cardinality of the set $\mathbb{T}$ i.e.,  $|\mathbb{T}|=d$. Then there exists a number  $p$ such that $p=q-2d$. Using it in Eq. (\ref{ajj'}), we get the observables for each Alice$_k$, and from Eq. (\ref{mnmi}), we find the observables of Bob which in turn fixes the state.
\section {Characterization of the non-$n$-local correlations}\label{VI}

We characterize the network nonlocality and show the correspondence with the standard Bell nonlocality. Such a characterization was first discussed by Gisin \emph{et. al.} \cite{Gisin2017} for the case of the bilocality scenario when all the parties receive two inputs.  They showed that any two-qubit quantum state that violates Clauser-Horne-Shimony-Halt  inequality also violates the bilocality inequality \cite{Cyril2012}. A similar correspondence is provided for the linear chain using two-qubit quantum states in \cite{kundu2020}. This characterization of network nonlocality was limited to the two-qubit systems only but in \cite{snehachsh},  we had also shown a correspondence between violations of various network inequalities and suitable Bell type inequalities without assuming any dimension of the system. 

The  characterization technique developed in \cite{snehachsh} is also applicable for the nonlocality of our arbitrary generalized $n$-locality scenario where Bob performs an arbitrary $m$ number of measurements.  Note that every source $S_k$ emits physical systems to  Alice$_k$ and Bob. In general, Alice$_k$ performs the measurement of the  observables $A^k_{x_k}$ upon receiving  input $x_k\in[2^{m-1}]$ and similarly Bob performs  measurement $B_i$ upon receiving  the input $i\in[m]$.  We use a suitable Bell-type inequality \cite{Ghorai2018} for Alice$_k$ and Bob and establish its one-one correspondence with the $n$-locality inequalities using the SOS approach. In quantum theory, the state $\rho_{A_kB}$ is shared between  Alice$_k$ and Bob. Then, Bob's observables can be taken as  $B_i=\bigotimes\limits_{k=1}^n B_i^k$,  by considering the $n$ number of subsystems received by Bob from each source $S_k$ with $k\in [n]$. 

As a case study, we first consider the scenario for $m=3$, i.e., Bob receives three inputs and each Alice receives four  inputs. For the inequality Eq. (\ref{deltan3}), it is already derived that 
$(\Delta^{n}_{3})_{Q}^{opt} =\sum\limits_{i=1}^{3}(\omega^{n}_{3,i})^{\frac{1}{n}}$
	where $\omega^{n}_{3,i}=\prod\limits_{k=1}^n(\omega^{n}_{3,i})_{A_{k}}$ and
\ba
\label{wn31ak}
&&(\omega^{n}_{3,1})_{A_{k}}=||A^{k}_{1}+A^{k}_{2}+A^{k}_{3}-A^{k}_{4}||_{2}\\
\nonumber&=&\bigg(4+\langle\{A^{k}_{1},(A^{k}_{2}+A^{k}_{3}-A^{k}_{4})\}+\{A^{k}_{2},(A^{k}_{3}-A^{k}_{4})\}-\{A^{k}_{3},A^{k}_{4}\}\rangle\bigg)^{1/2}\ea Similarly,  we can write for
	$(\omega^{n}_{3,2})_{A_{k}}$ and $ (\omega^{n}_{3,3})_{A_{k}}$, $\forall k\in[n]$. From the  inequality  Eq. (\ref{Tavakoli}) by using SOS approach,  we get that \ba\label{wn3i2}(\Delta^{n}_{3})_{Q}^{opt}=\sum\limits_{i=1}^{3}(\omega^{n}_{3,i})^{\frac{1}{n}}\leq\prod\limits_{k=1}^{n}\bigg( \sum\limits_{i=1}^{3}(\omega^{n}_{3,i})_{A_{k}}\bigg)^{\frac{1}{n}}\ea 
	
Now,	we consider that the $k^{th}$ source emitting physical system to Alice$_{k}$ and Bob, we construct a suitable two-party Bell  inequality as  \ba\nonumber \label{Bk3}\mathcal{B}^k_{3}&=&(A^{k}_{1}+A^{k}_{2}+A^{k}_{3}-A^{k}_{4}) B_{1}+(A^{k}_{1}+A^{k}_{2}-A^{k}_{3}+A^{k}_{4}) B_{2}\\&+&(A^{k}_{1}-A^{k}_{2}+A^{k}_{3}-A^{k}_{4}) B_{3}\leq 4\ea
 where the observables of each Alice$_k$ is bounded by the following condition:
 \ba A^k_1-A^k_2-A^k_3-A^k_4=0.\ea
		We first derive the optimal quantum value of  $(\mathcal{B}^k_{3})_{Q}$ using SOS approach. Following the steps used earlier, we show that there is a positive semidefinite operator $\langle\epsilon_{3}\rangle\geq 0$, that can be expressed as $\langle \epsilon_{3}\rangle_{Q}=-(\mathcal{B}^k_{3})_{Q}+\beta^k_{3}$. Here $\beta^k_{3}$ is the optimal value and  can be obtained when $\langle \epsilon^k_{3}\rangle_{Q}$ is equal to zero. To prove this, let us  consider a set of suitable positive operators $D^k_{3,i}$ which is polynomial functions of $A^{k}_{x_{k}}, (\forall k)$ and $B_{i}$  so that 
	\begin{eqnarray}
	\label{gammak3}
 \nonumber
\langle\epsilon_{3}\rangle&=&\dfrac{(\omega_{3,1}^{n})_{A_k}}{2}\langle\psi|(D^k_{3,1})^{\dagger}(D^k_{3,1})|\psi\rangle+ \frac{(\omega_{3,2}^{n})_{A_k}}{2}\langle\psi|(D^k_{3,2})^{\dagger}(D^k_{3,2})|\psi\rangle\\&&+\dfrac{(\omega_{3,3}^{n})_{A_k}}{2}\langle\psi|(D^k_{3,3})^{\dagger}(D^k_{3,3}|\psi\rangle\end{eqnarray}
	where $(\omega^{n}_{3,i})_{A_k}$  is already  defined above.
	We choose a  suitable set of  positive operators $D^k_{3,i}$s such that 
\ba\nonumber
    D^k_{3,1}|\psi\rangle&=&\left(\frac{{A}^{k}_{1}+{A}^{k}_{2}+A^{k}_{3}-A^{k}_{4}}{(\omega_{3,1}^{n})_{A_{k}}}-B_{1}\right) |\psi\rangle\\
D^k_{3,2}|\psi\rangle&=&\left(\frac{{A}^{k}_{1}+{A}^{k}_{2}-A^{k}_{3}+A^{k}_{4}}{(\omega_{3,2}^{n})_{A_{k}}}-B_2\right)|\psi\rangle\\\nonumber D^k_{3,3}|\psi\rangle&=&\left(\frac{{A}^{k}_{1}-{A}^{k}_{2}+A^{k}_{3}+A^{k}_{4}}{(\omega_{3,3}^{n})_{A_{k}}}-B_3\right) |\psi\rangle\hspace{8pt}
\ea
Substituting these in Eq. (\ref{gammak3}), we get
	\ba\langle\epsilon_{3}\rangle_{Q}=-(\mathcal{B}^k_{3})_{Q} +\left[(\omega^{n}_{3,1})_{A_{k}}+(\omega^{n}_{3,2})_{A_{k}}+(\omega^{n}_{3,3})_{A_{k}}\right]\hspace{15pt}\ea 	The optimal value of $(\mathcal{B}^{k}_{3})_{Q}$ is obtained if  $\langle \gamma^k_{3}\rangle_{Q}=0$.	Hence,
	\begin{eqnarray}(\mathcal{B}^k_{3})_{Q}^{opt} =(\omega^{n}_{3,1})_{A_{k}}+(\omega^{n}_{3,2})_{A_{k}}+(\omega^{n}_{3,3})_{A_{k}}
	\end{eqnarray}
 From Eq. (\ref{wn3i2}), we can write
 \ba(\Delta^n _{3})_{Q}^{opt}\leq\prod\limits_{k=1}^n[(\mathcal{B}^k_{3})_{Q}^{opt}]^{1/n} \ea
 This implies that if for each $k\in[n]$, if a two-party quantum state $\rho_{A_kB}$ violates the inequality $ \mathcal{B}^k_3\leq 4$ then $\bigotimes\limits_{k=1}^n\rho_{A_kB}$ must violate the $n$-locality  inequality in Eq. (\ref{deltan3}). 
 
 This characterization can be extended upto the generalized $n$-locality scenario with arbitrary $m$ number of inputs for Bob.  For the generalized  inequality in Eq. (\ref{deltapncnm}), we  have already derived 	\begin{align}
	\label{optbnm2}
	(\Delta^{n}_{m})_{Q}^{opt} &= \sum\limits_{i=1}^{m}(\omega^{n}_{m,i})^{\frac{1}{n}}
\leq \prod \limits_{k=1}^{n} \bigg(\sum\limits_{i=1}^{m}(\omega^{n}_{m,i})_{A_k}\bigg)^{\frac{1}{n}}\end{align}
where $(\omega^{n}_{m,i})_{A_k}$ is defined already in Eq. (\ref{wn31ak}).  We consider that the source $S_{k}$ emits a quantum state $\rho_{A_kB}$  which is shared by  Alice$_{k}$ and Bob. By considering the observables of Alice$_k$ and Bob are given by $A^{k}_{x_k}$, ($x_k\in[2^{m-1}]$) and $B_{i}$, ($i\in[m]$) respectively such that the observables of Alice$_k$  are constrained by the equations Eq. (\ref{conditionm})
we can use  a suitable two-party Bell inequality \cite{Ghorai2018} as 
\ba\label{Bkm}\mathcal{B}^k_m=\sum\limits_{i=1}^{m}\sum\limits_{x_{k}=1}^{2^{m-1}}(-1)^{y^{x_{k}}_{i}} A_{x_{k}}^{k}B_{i}\leq 2^{m-1}\ea
To formulate this inequality, a parallel encoding scheme needs to be  used, as explicitly stated in Sec.\ref{V}. Following a similar SOS approach, we get  
\ba
\label{hh}(\mathcal{B}^k_m)_Q^{opt}=\sum\limits_{i=1}^{m}(\omega^{n}_{m,i})_{A_k}\ea
 Hence, substituting Eq. (\ref{hh}) in Eq. (\ref{optbnm2}), we finally get that 	\ba(\Delta^{n}_{m})_{Q}^{opt} 
\leq \prod \limits_{k=1}^{n}(\mathcal{B}^k_m)_Q^{opt}  \ea
which implies that for each $k\in[n]$, if the state  $\rho_{A_kB}$ violates the inequality (\ref{Bkm}), then $\bigotimes\limits_{k=1}^n\rho_{A_kB}$ must violate the generalized $n$-locality inequality in Eq.  (\ref{deltapncnm}). Importantly, the one-to-one  correspondence of the quantum violation of $n$-locality inequality and a suitable Bell nonlocality is demonstrated  without assuming the dimension of the quantum system.

\section{Summary and Discussion}
\label{VII}
In summary, we considered the star-network configuration that features $n$ independent sources.   The network involves a central party Bob, encircled by  $n$ number of edge parties, denoted by Alice$_{k}$, $k\in[n]$. Each edge party Alice$_{k}$ shares a physical system with the central party Bob, generated by an independent source. Based on the $n$-locality assumption corresponding to the independent sources, we formulated a generalized form of $n$-locality  inequality in an arbitrary $m$  input scenario. This is in contrast to the earlier works that considered only two inputs for each party.  In our arbitrary input quantum network, Bob performs the measurement of $m$ dichotomic observables, and each Alice$_{k}$ performs $2^{m-1}$ dichotomic measurements. All measurements produce binary outputs. We consider that the joint probabilities are bounded by certain linear conditions  that impose constraints on the observables of each edge party. 

For the sake of a better understanding of our generalized results, we first demonstrated the quantum violation of  $n$-locality inequality for a three-input ($m=3$) scenario where  Bob performs three binary-outcome measurements and each of the  Alice$_{k}$s ($k\in[n]$) performs four binary-outcome measurements. We introduced an elegant SOS approach to derive the optimal quantum violation.  We explicitly derived the observables required for Alice and Bob as well the state for obtaining the optimal quantum value. Importantly, the derivation of optimal quantum violation is independent of the dimension of the system.  

Further, we proposed $n$-locality  inequalities for   $m=4$ and $m=5$ and provided explicit derivation of optimal quantum violations again without assuming the dimension of the system. We generalized our treatment where Bob receives arbitrary $m$ input and each Alice receives $2^{m-1}$ inputs. Using the aforementioned  SOS approach, we optimized the quantum violation of the generalized $n$-locality inequality along with the constraints that must be satisfied by the observables of each Alice$_k$. This indeed certifies the conditional dependence of the observables of each edge party. Importantly, unlike the earlier works, our derivation is without assuming the dimension of the system.   We demonstrated that to obtain the optimal quantum violation, each Alice has to share  $\lfloor\frac{m}{2}\rfloor$ copies of two-qubit maximally entangled states with Bob.

Furthermore, we have characterized the network nonlocality by demonstrating a one-to-one correspondence between the optimal quantum violations of the generalized $n$-locality inequality and the suitable bipartite Bell inequality \cite{Ghorai2018}. This implies that if we consider $\rho_{A_kB}$ is the state that is  generated by the source $S_k$ and shared between Alice$_k$ and Bob then this characterization implies that  if for each $k\in[n]$, $\rho_{A_kB}$ violates suitable bipartite Bell  inequality \cite{Ghorai2018} then $\bigotimes\limits_{k=1}^{n}\rho_{A_kB}$ violates the   generalized $n$-locality inequality when Alice$_k$ and Bob  performs $2^{m-1}$ and $m$ number of dichotomic measurements respectively.

 We note here that in one of our earlier works \cite{Sneha2021} we considered that the central party Bob receives  $2^{m-1}$ inputs, and each Alice receives  $m$ inputs. In our present work, the number of inputs is swapped for Bob and each Alice, and hence the present work has a structural resemblance to our previous work\cite{Sneha2021}. However, we argue that such a resemblance is structural only.  We note the fact that if we would consider the simple bipartite scenario involving only Alice and Bob, then such swapping of the number of measurement settings would have been trivial. But, in a network scenario, the situation is not so straightforward. This is due to the complexity of the form of the inequalities, which involve the nonlinear terms, the modulus, and the exponents.  Most importantly, here we impose  additional conditions on the inputs of each party.  As a result, the $n$-locality inequality and the optimization of quantum value becomes completely  non-trivial even if we are  swapping the number of measurements between central and edge parties.  More over the dimension independent optimization of quantum value itself certifies the conditional dependence of the inputs of each edge party and hence , the set of observables required for obtaining the optimal quantum value in our present work is completely different from the set of observables required in \cite{Sneha2021}.  For example, in \cite{Sneha2021} Alice's  observables  require to be anti-commuting for any $n$, but here Alice's  observables are neither commuting nor anticommuting. Rather the optimal quatum violation certifies some typical conditional dependence of the observables. This establishes the fact that although there is a structural resemblance between the two scenarios, the $n$-locality inequalities are 
 non-identical and the optimal quantum values self-test different sets of observables.

We conclude by pointing out a few interesting directions for future research. There could be one more implication towards providing  device-independent security in quantum network.  Recently \cite{cab21}, it is shown that if a   device-independent
quantum key distribution protocol is based on a Bell’s inequality, then it  works better for lower detection efficiency, when  considered with the number of inputs are more than  two. Hence, here also, it will be exciting to see if  our scheme of   $n$-locality inequalities with arbitrary $m$ number of inputs can  enhance the security of the protocol across the network. We note again that the elegant SOS approach used in this work plays a crucial role in dimension-independent optimization and characterization of network nonlocality. Such a feature may be useful to formulate the device-independent self-testing of states and measurements based on the optimal quantum violation of $n$-locality inequality.  However, the correlations in a quantum network are not convex, and this dimension-independent optimization may enable one to construct device-independent certification protocols in the quantum network, as initiated in \cite{Kerstjens2019}. This could also be an interesting line of study.


  
 \section{Acknowledgment}SM acknowledges the support from the research grant DST/ICPS/QuEST/2019/4. AKP acknowledges the support from the research grant MTR/2021/000908.
\appendix
\begin{widetext}
	\section{Detailed derivation of optimal quantum violation of $n$-locality inequality for $m=4$}
	\label{A}
In this case each of the $n$ Alices performs the measurements of eight binary-outcome observables and the central party Bob performs four measurements. Alice$_{k}(k\in[n])$ considers eight dichotomic  measurements $A^{k}_{x_{k}}$ with $x_{k}\in[8]$. Bob performs  the measurements of  four dichotomic observables  $B_{i}$ with $i=1,2,3,4$ on the $n$ sub-systems he receives from $n$-number of independent sources $S_{k}$. The  $n$-locality inequality we proposed in the main text in Eq. (\ref{deltapncn4}) is given by
	\begin{equation}
	\label{ADelta4}
	(\Delta^{n}_{4})_{nl}\leq |I^{n}_{4,1}|^{\frac{1}{n}}+|I^{n}_{4,2}|^{\frac{1}{n}}+|I^{n}_{4,3}|^{\frac{1}{n}}+|I^{n}_{4,4}|^{\frac{1}{n}}\leq 8
	\end{equation}

	As already mentioned, we derive the optimal quantum value of $(\Delta^{n}_{4})_{Q}$ by using  an elegant version of SOS approach. Following the technique developed in Sec. \ref{III}  for $m=3$, we show that there is a positive semidefinite operator $\langle\gamma^{n}_{4}\rangle\geq 0$, that can be expressed as $\langle \gamma^n_{4}\rangle_{Q}=-(\Delta^{n}_{4})_{Q}+\beta^{n}_{4}$. Here $\beta^{n}_{4}$ is the optimal value that can be obtained when $\langle \gamma^n_{4}\rangle_{Q}$ is equal to zero. To  prove this, we  consider a set of suitable positive operators $M^{n}_{4,i}$ which is polynomial functions of   $A^{k}_{x_{k}}$, $B_{i}$ such that 
	\begin{equation}
	\label{gamma4}
	\langle\gamma^{n}_{4}\rangle=\dfrac{(\omega_{4,1}^{n})^{\frac{1}{n}}}{2}|(M^{n}_{4,1})|\psi\rangle|^2+ \frac{(\omega_{4,2}^{n})^{\frac{1}{n}}}{2}|(M^{n}_{4,2})\psi\rangle|^2+\dfrac{(\omega_{4,3}^{n})^{\frac{1}{n}}}{2}|(M^{n}_{4,3}|\psi\rangle|^2+\dfrac{(\omega_{4,4}^{n})^{\frac{1}{n}}}{2}|(M^{n}_{4,4}|\psi\rangle|^2\end{equation}
	where $\omega^{n}_{4,i}$ is suitable positive numbers and $\omega^{n}_{4,i}=\prod\limits_{k=1}^n(\omega^{n}_{4,i})_{A_{k}}$. The optimal quantum value of $(\Delta^{n}_{4})_{Q}$ is obtained if $\langle \gamma^{n}_{4}\rangle_{Q}=0$, implying that	$|M^{n}_{4,i}|\psi\rangle|=0$ (for notational convenience, we use $|\psi\rangle_{A_1A_2\cdots A_nB}=|\psi\rangle$). We choose a  suitable set of  positive numbers $|M^{n}_{4,i}|\psi\rangle|$, given by
	
\ba
\label{mn41}
    \nonumber
|M^{n}_{4,1}|\psi\rangle|=\bigg|\prod\limits_{k=1}^{n}\left(\frac{A^{k}_{1}+A^{k}_{2}+A^{k
		}_{3}+A^{k}_{4}-A^{k}_{5}+A^{k
		}_{6}+A^{k}_{7}+A^{k}_{8}}{(\omega_{4,1}^{n})_{A_{k}}}\right)|\psi\rangle\bigg|^{\frac{1}{n}}
		-|B_{1}|\psi\rangle|^{\frac{1}{n}}\\
		  |M^{n}_{4,2}|\psi\rangle|=\bigg|\prod\limits_{k=1}^{n}\left(\frac{A^{k}_{1}+A^{k}_{2}+A^{k
		}_{3}-A^{k}_{4}+A^{k}_{5}+A^{k
		}_{6}-A^{k}_{7}-A^{k}_{8}}{(\omega_{4,2}^{n})_{A_{k}}}\right)|\psi\rangle\bigg|^{\frac{1}{n}}
		-|B_{2}|\psi\rangle|^{\frac{1}{n}}\\
	\nonumber
 |M^{n}_{4,3}|\psi\rangle|=\bigg|\prod\limits_{k=1}^{n}\left(\frac{{A}^{k}_{1}+{A}^{k}_{2}-A^{k}_{3}+A^{k}_{4}+A^{k}_{5}-A^{k
		}_{6}+A^{k}_{7}-A^{k}_{8}}{(\omega_{4,3}^{n})_{A_{k}}}\right)|\psi\rangle\bigg|^{\frac{1}{n}}
		-|B_{3}|\psi\rangle|^{\frac{1}{n}}\\
	\nonumber
	 |M^{n}_{4,4}|\psi\rangle|=\bigg|\prod\limits_{k=1}^{n}\left(\frac{{A}^{k}_{1}-{A}^{k}_{2}+A^{k}_{3}+A^{k}_{4}+A^{k}_{5}-A^{k
		}_{6}-A^{k}_{7}+A^{k}_{8}}{(\omega_{4,4}^{n})_{A_{k}}}\right)|\psi\rangle\bigg|^{\frac{1}{n}}
		-|B_{4}|\psi\rangle|^{\frac{1}{n}}
	\ea	

 Following the argument stated in Sec \ref{III}, we can show that $|M^{n}_{4,i}|\psi\rangle|$ is indeed a positive number.
	Using these in Eq. (\ref{gamma4}), we get 
	$\langle\gamma^{n}_{4}\rangle_{Q}=-(\Delta^{n}_{4})_{Q} +\sum\limits_{i=1}^{4}\left(\omega^{n}_{4,1}\right)^{\frac{1}{n}}$. 	The optimal value of $(\Delta^{n}_{4})_{Q}$ is obtained if  $\langle \gamma^n_{4}\rangle_{Q}=0$.	Hence,
	\begin{eqnarray}(\Delta^{n}_{4})_{Q}^{opt} =\sum\limits_{i=1}^{4}(\omega^{n}_{4,i})^{\frac{1}{n}}\end{eqnarray}
	where $(\omega^{n}_{4,1})_{A_{k}}$ is defined as 
	\ba\nonumber
	(\omega^{n}_{4,1})_{A_{k}}&=&|| A^{k}_{1}+A^{k}_{2}+A^{k
		}_{3}+A^{k}_{4}-A^{k}_{5}+A^{k
		}_{6}+A^{k}_{7}+A^{k}_{8}||_{2}\\\nonumber
		&=&[8+\langle\psi|\{A^{k}_{1},(A^{k}_{2}+A^{k
		}_{3}+A^{k}_{4}-A^{k}_{5}+A^{k
		}_{6}+A^{k}_{7}+A^{k}_{8})\}+\{A^{k}_{2},(A^{k
		}_{3}+A^{k}_{4}-A^{k}_{5}+A^{k
		}_{6}+A^{k}_{7}+A^{k}_{8})\}+\{A^{k}_{3},(A^{k}_{4}-A^{k}_{5}+A^{k
		}_{6}+A^{k}_{7}\\&&+A^{k}_{8}\}+
		\{A^{k}_{4},(-A^{k}_{5}+A^{k
		}_{6}+A^{k}_{7}+A^{k}_{8}\}-\{A^{k}_{5},(A^{k
		}_{6}+A^{k}_{7}+A^{k}_{8}\}+\{A^{k}_{6},(A^{k}_{7}+A^{k}_{8}\}+\{A^{k}_{7},(A^{k}_{8}\}|\psi\rangle]^{\frac{1}{2}}\ea
  
  Similarly,  we can write for $(\omega^{n}_{4,2})_{A_{k}}=||A^{k}_{1}+A^{k}_{2}+A^{k
		}_{3}-A^{k}_{4}+A^{k}_{5}+A^{k
		}_{6}-A^{k}_{7}-A^{k}_{8}||_{2}$, \hspace{5pt}$
		(\omega^{n}_{4,3})_{A_{k}}=||A^{k}_{1}+A^{k}_{2}-A^{k
		}_{3}+A^{k}_{4}+A^{k}_{5}-A^{k
		}_{6}+A^{k}_{7}-A^{k}_{8}||_{2}$,  $(\omega^{n}_{4,4})_{A_{k}}=||A^{k}_{1}-A^{k}_{2}+A^{k
		}_{3}+A^{k}_{4}+A^{k}_{5}-A^{k
		}_{6}-A^{k}_{7}+A^{k}_{8}||_{2}.$ Since  $\omega^{n}_{4,i}=\prod\limits_{k=1}^n(\omega^{n}_{4,i})_{A_k}, i=1,2,3,4$, by using the  inequality  (\ref{Tavakoli}), we get  \ba\sum\limits_{i=1}^{4}(\omega^{n}_{4,i})^{\frac{1}{n}}\leq \prod\limits_{k=1}^n\left(\sum\limits_{i=1}^{4}(\omega^{n}_{4,i})_{A_{k}}\right)^\fn\ea
	Using convex inequality,  \ba\sum\limits_{i=1}^{4}(\omega^{n}_{4,i})_{A_{k}}\leq \sqrt{4\sum\limits_{i=1}^{4}\bigg((\omega^{2}_{4,i})_{A_{k}}\bigg)^2},\quad  \forall k\in[n]\ea and noting that each  observable $A_{x_k}^k$ is dichotomic, we can write $\sum\limits_{i=1}^{4}\bigg((\omega^{2}_{4,i})_{A_{k}}\bigg)^2$ as following: 
	\begin{eqnarray}
		\label{w24ik}\nonumber
	\sum\limits_{i=1}^{4}\bigg((\omega^{2}_{4,i})_{A_{k}}\bigg)^2&=& \langle\psi|32+2[\{A^{k}_{1},(A^{k}_{2}+A^{k}_{3}+A^{k}_{4}+A^{k}_{5})\}+\{A^{k}_{2},(A^{k}_{6}+A^{k}_{7}-A^{k}_{8})\}+\{A^{k}_{3},(A^{k}_{6}-A^{k}_{7}+A^{k}_{8})\}+\{A^{k}_{4},(-A^{k}_{6}+A^{k}_{7}+A^{k}_{8})\}\\&&-\{A^{k}_{5},(A^{k}_{6}+A^{k}_{7}+A^{k}_{8})]\}|\psi\rangle=\langle\psi|(32+\delta_4)|\psi\rangle
		\end{eqnarray}\\
		where 
  \ba\na\delta_4=2\bigg[\{A^{k}_{1},(A^{k}_{2}+A^{k}_{3}+A^{k}_{4}+A^{k}_{5})\}+\{A^{k}_{2},(A^{k}_{6}+A^{k}_{7}-A^{k}_{8})\}+\{A^{k}_{3},(A^{k}_{6}-A^{k}_{7}+A^{k}_{8})\}+\{A^{k}_{4},(-A^{k}_{6}+A^{k}_{7}+A^{k}_{8})\}-\{A^{k}_{5},(A^{k}_{6}+A^{k}_{7}+A^{k}_{8})\bigg]\\\ea
		Let us consider the following states \begin{eqnarray}
		|\psi_{1}\rangle&=&(A^{k}_{1}+A^{k}_{2}-A^{k
		}_{3}-A^{k}_{4}-A^{k}_{5}-A^{k
		}_{6}-A^{k}_{7}+A^{k}_{8})|\psi\rangle; \hspace{6pt}
		|\psi_{2}\rangle=(A^{k}_{1}-A^{k}_{2}+A^{k
		}_{3}-A^{k}_{4}-A^{k}_{5}-A^{k
		}_{6}+A^{k}_{7}-A^{k}_{8})|\psi\rangle;\\
		\nonumber
		|\psi_{3}\rangle&=&(A^{k}_{1}-A^{k}_{2}-A^{k
		}_{3}+A^{k}_{4}-A^{k}_{5}+A^{k
		}_{6}-A^{k}_{7}-A^{k}_{8})|\psi\rangle; \hspace{6pt}
		|\psi_{4}\rangle=(A^{k}_{1}-A^{k}_{2}-A^{k
		}_{3}-A^{k}_{4}+A^{k}_{5}+A^{k
		}_{6}+A^{k}_{7}+A^{k}_{8})|\psi\rangle
		\end{eqnarray}  such that $|\psi\rangle\neq 0$. Then
		\begin{eqnarray}\label{deltai}
		    \langle\psi_{1}|\psi_{1}\rangle=\langle\psi|(8+(\delta_4)_{1})|\psi\rangle \implies \langle(\delta_4)_{1}\rangle=8-\langle\psi_{1}|\psi_{1}\rangle, \hspace{3pt}
		    \langle\psi_{2}|\psi_{2}\rangle=\langle\psi|(8+(\delta_4)_{2})|\psi\rangle \implies \langle(\delta_4)_{2}\rangle=8-\langle\psi_{2}|\psi_{2}\rangle\\
		    \nonumber
		    \langle\psi_{3}|\psi_{3}\rangle=\langle\psi|(8+(\delta_4)_{3})|\psi\rangle \implies \langle(\delta_4)_{3}\rangle=8-\langle\psi_{3}|\psi_{3}\rangle, \hspace{3pt}
		    \langle\psi_{4}|\psi_{4}\rangle=\langle\psi|(8+(\delta_4)_{4})|\psi\rangle \implies \langle(\delta_4)_{4}\rangle=8-\langle\psi_{4}|\psi_{4}\rangle
		    \end{eqnarray}such that $\sum\limits_{l=1}^{4}\langle(\delta_4)_{l}\rangle=\langle\delta\rangle$. 
	Clearly, adding Eqs. of (\ref{deltai}), we get $\langle\delta_4\rangle=32-(\langle\psi_{1}|\psi_{1}\rangle+\langle\psi_{2}|\psi_{2}\rangle+\langle\psi_{2}|\psi_{2}\rangle+\langle\psi_{2}|\psi_{2}\rangle)$. Hence,  $\langle\delta_4\rangle_{max}$ is obtained when $\langle\psi_{i}|\psi_{i}\rangle=0$, $ \forall i \in[4]$. Since, $|\psi\rangle\neq 0$, then 	\begin{eqnarray}\label{po4}
		A^{k}_{1}+A^{k}_{2}-A^{k
		}_{3}-A^{k}_{4}-A^{k}_{5}-A^{k
		}_{6}-A^{k}_{7}+A^{k}_{8}&=&0\\
		\nonumber
		A^{k}_{1}-A^{k}_{2}+A^{k
		}_{3}-A^{k}_{4}-A^{k}_{5}-A^{k
		}_{6}+A^{k}_{7}-A^{k}_{8}&=&0\\
		\nonumber
		A^{k}_{1}-A^{k}_{2}-A^{k
		}_{3}+A^{k}_{4}-A^{k}_{5}+A^{k
		}_{6}-A^{k}_{7}-A^{k}_{8}&=&0\\
		\nonumber
		A^{k}_{1}-A^{k}_{2}-A^{k
		}_{3}-A^{k}_{4}+A^{k}_{5}+A^{k
		}_{6}+A^{k}_{7}+A^{k}_{8}&=&0
		\end{eqnarray}
Hence, to obtain the optimal value of $(\Delta^{n}_{4})_{Q}^{opt}$, the  conditions given in Eq.(\ref{po4}) have to be satisfied. Using them, we get  $\langle\delta_4\rangle_{max}=32$. Thus from Eq. (\ref{w24ik}), we get $\sum\limits_{i=1}^{4}\bigg((\omega^{2}_{4,i})_{A_{k}}\bigg)^2=64$. Plugging it in the above-mentioned convex inequality, finally we get $\sum\limits_{i=1}^{4}(\omega^{n}_{4,i})^\fn\leq \prod\limits_{k=1}^n\bigg(\sum\limits_{i=1}^{4}(\omega^{n}_{4,i})_{A_{k}}\bigg)^\fn\leq16$ i.e., \ba (\Delta^{n}_{4})_{Q}^{opt}= 16 .\ea
 Since each observable $A^k_{x_k}$ is dichotomic, pre-multiplying and post-multiplying the first equation of Eq.(\ref{po4}) by $A^k_1$, and adding them we get
	 
	 	\ba\label{AK4}-\{A^k_1,A^k_2\}+\{A^k_1,A^k_3\}+\{A^k_1,A^k_4\}+\{A^k_1,A^k_5\}+\{A^k_1,A^k_6\}+\{A^k_1,A^k_7\}-\{A^k_1,A^k_8\}=2\ea

	  Similarly,  other $31$ relations can be found following the  similar method.   Solving these equations, we get  that Alice's observables has to satisfy the following relations to obtain the optimal quantum value. \ba\label{aanti4}&\{A^{k}_{1}, A^{k}_{2}\}&=\{A^{k}_{1}, A^{k}_{3}\}=\{A^{k}_{1}, A^{k}_{4}\}=\{A^{k}_{1}, A^{k}_{5}\}=\{A^{k}_{2}, A^{k}_{6}\}=\{A^{k}_{2}, A^{k}_{7}\}=\{A^{k}_{3}, A^{k}_{6}\}= \{A^{k}_{3}, A^{k}_{8}\}= \{A^{k}_{4}, A^{k}_{7}\}= \{A^{k}_{4}, A^{k}_{8}\}=1,\\\nonumber
	&\{A^{k}_{1}, A^{k}_{6}\}&=\{A^{k}_{1}, A^{k}_{7}\}=\{A^{k}_{1}, A^{k}_{8}\}=\{A^{k}_{2}, A^{k}_{3}\}=\{A^{k}_{2}, A^{k}_{4}\}=\{A^{k}_{2}, A^{k}_{5}\}=\{A^{k}_{3}, A^{k}_{4}\}=\{A^{k}_{3}, A^{k}_{5}\}=\{A^{k}_{4}, A^{k}_{5}\}=\{A^{k}_{6}, A^{k}_{7}\}= \{A^{k}_{6}, A^{k}_{8}\}=\{A^{k}_{7}, A^{k}_{8}\}=0,\\\nonumber
	&\{A^{k}_{2}, A^{k}_{8}\}&=\{A^{k}_{3}, A^{k}_{7}\}=\{A^{k}_{4}, A^{k}_{6}\}=\{A^{k}_{5}, A^{k}_{6}\}= \{A^{k}_{5}, A^{k}_{7}\}=\{A^{k}_{5}, A^{k}_{8}\}=-1.   
	\ea
Next, we 
Bob's observables can be found from the optimization conditions i.e.,  for each $i\in[4]$, 	$|M^{n}_{4,i}|\psi\rangle|=0$. This, in turn, fixes the required state, which has to be at least two copies of two-qubit maximally entangled states shared by each Alice$_{k}$ and Bob.  
	\section{Detailed derivation of optimal quantum violation of $n$-locality inequality for $m=5$}
	 	\label{B}
In this case each of the $n$ Alices performs the measurements of sixteen  binary-outcome observables and the central party Bob performs five measurements. Alice$_{k}(k\in[n])$ considers eight dichotomic  measurements $A^{k}_{x_{k}}$ with $x_{k}\in[16]$, $k\in[n]$ and Bob performs  the measurements of  five dichotomic observables  $B_{i}$ with $i=1,2,3,4,5$ on the $n$ sub-systems he receives from $n$-number of independent sources $S_{k}$.

The  $n$-locality inequality we proposed in the main text in Eq. (\ref{deltapncn4}) is given by
	\begin{equation}
	\label{Delta5}
	(\Delta^{n}_{5})_{nl}\leq \sum\limits_{i=1}^{5}|I^{n}_{5,i}|^\fn \leq 16
	\end{equation} where $I^{n}_{5,i}$ s $(i \in[8])$ are defined as linear combinations of correlations are such as
	\begin{eqnarray}
		\nonumber
		&& I_{5,1}^{n}=\langle\prod\limits_{k=1}^{n}(A^{k}_{1}+A^{k}_{2}+A^{k
		}_{3}+A^{k}_{4}+A^{k}_{5}-A^{k
		}_{6}+A^{k}_{7}+A^{k}_{8}+A^{k}_{9}+A^{k}_{10}+A^{k
		}_{11}+A^{k}_{12}-A^{k}_{13}-A^{k
		}_{14}-A^{k}_{15}-A^{k}_{16})B_{1}\rangle\\
		\nonumber
		&& I_{5,2}^{n}=\langle\prod\limits_{k=1}^{n}(A^{k}_{1}+A^{k}_{2}+A^{k
		}_{3}+A^{k}_{4}-A^{k}_{5}+A^{k
		}_{6}+A^{k}_{7}+A^{k}_{8}+A^{k}_{9}-A^{k}_{10}-A^{k
		}_{11}-A^{k}_{12}+A^{k}_{13}+A^{k
		}_{14}+A^{k}_{15}-A^{k}_{16})B_{2}\rangle\\
		&&I_{5,3}^{n}=\langle\prod\limits_{k=1}^{n}(A^{k}_{1}+A^{k}_{2}+A^{k
		}_{3}-A^{k}_{4}+A^{k}_{5}+A^{k
		}_{6}+A^{k}_{7}-A^{k}_{8}-A^{k}_{9}+A^{k}_{10}+A^{k
		}_{11}-A^{k}_{12}+A^{k}_{13}+A^{k
		}_{14}-A^{k}_{15}+A^{k}_{16})B_{3}\rangle\hspace{10pt}\\ \nonumber
		&& I_{5,4}^{n}=\langle\prod\limits_{k=1}^{n}(A^{k}_{1}+A^{k}_{2}-A^{k
		}_{3}+A^{k}_{4}+A^{k}_{5}+A^{k
		}_{6}-A^{k}_{7}+A^{k}_{8}-A^{k}_{9}+A^{k}_{10}-A^{k
		}_{11}+A^{k}_{12}+A^{k}_{13}-A^{k
		}_{14}+A^{k}_{15}+A^{k}_{16})B_{4}\rangle\hspace{10pt}\\ \nonumber
		&&I_{5,5}^{n}=\langle\prod\limits_{k=1}^{n}(A^{k}_{1}-A^{k}_{2}+A^{k
		}_{3}+A^{k}_{4}+A^{k}_{5}+A^{k
		}_{6}-A^{k}_{7}-A^{k}_{8}+A^{k}_{9}-A^{k}_{10}+A^{k
		}_{11}+A^{k}_{12}-A^{k}_{13}+A^{k
		}_{14}+A^{k}_{15}+A^{k}_{16})B_{5}\rangle
		\end{eqnarray}
Now we  consider the set $\mathcal{L}_5=\{s|s\in\{0,1\}^5, \sum_r s_r\geq 2\}$, $r\in\{1,2,\cdots 5\}$. The element  $s_l\in \mathcal{L}_5$ is  such that $(s_l)_r\neq 2u$, for some $u\in\mathbb{N}$. We then find $11$  such elements  $s_l$  where $l\in[11]$. 		
Let us assume that the joint  probabilities are constrained by the following conditions: 
\ba\label{condition5prob}
\sum\limits_{x_k\in U_{k,l}} P(a_k=0,b,\alpha_k|x_k,i,\chi_k)+\sum\limits_{x_k\notin U_{k,l}} P(a_k=1,b,\alpha_k|x_k,i,\chi_k)=
\sum\limits_{x_k\in U_{k,l}} P(a_k=1,b,\alpha_k|x_k,i,\chi_k)+\sum\limits_{x_k\notin U_{k,l}} P(a_k=0,b,\alpha_k|x_k,i,\chi_k)\hspace{3mm}\ea
 where we define the set $U_{k,l}\subset [16]$ as the collection of $x_k$s  such that for a given $s_l$,  $(-1)^{s_l.y^{x_k}}=1$ i.e., 
 \ba U_{k,l}=\{x_k:(-1)^{s_l.y^{x_k}}=1\}\ea Also for notational convenience, we denote that the set $\alpha_k$ as the following collection of elements   $\alpha_k=\{a_1,a_2,\dots, a_{k-1},a_{k+1},\dots, a_n\}$ and $\chi_k=\{x_1,x_2,\dots, x_{k-1},x_{k+1},\dots, x_n\}$ and it holds $\forall i\in[m], k\in[n]$.
From Eq. (\ref{conditionmprob}), we get that the observables  for each edge party  Alice$_k$  constrained by  the following $11$ conditions:
\begin{eqnarray}\label{condition5}
\sum\limits_{x_{k}=1}^{2^{m-1}} (-1)^{s_l.y^{x_{k}}}A^k_{x_k}=0, \forall s_l\in[2^{m-1}-m],k\in[n].\ea

	Let us first prove the inequality in Eq. (\ref{Delta5}) here. Here each Alice$_k$ shares the state $\lambda_{k}$ with Bob which is originated from the source $S_{k}$. In $n$-locality scenario, it is assumed that the sources are independent that they have no prior correlation. Hence the joint distribution of $\lambda_{k}$s ($k\in[n])$ given by $\rho(\lambda_{1}, \lambda_{2},\cdots\lambda_{n} )$ can be factorized as  $\rho(\lambda_{1}, \lambda_{2},\cdots\lambda_{n} )=\prod\limits_{k=1}^{n}\rho_k(\lambda_{k})$. Let us define  $\langle{A^{k}_{x_{k}}}\rangle_{\lambda_{k}} = \sum_{a_{k}}(-1)^{a_{k}}  P(a_{k}|x_{k},{\lambda_{k}})$, (similar definition holds for $\langle{B_{i}}\rangle_{\lambda_{1},\lambda_{2},\cdots\lambda_{n}}$). 
	Since,  $|\langle{B_{1}}\rangle_{\lambda_{1},\lambda_{2},\cdots\lambda_{n}}|\leq{1},$ and using the  independence assumption we can write
	\begin{eqnarray}|	I^{n}_{5,1}|\leq&\prod\limits_{k=1}^{n}\bigg|&\langle{A^{k}_{1}}\rangle_{\lambda_{k}}+\langle{A^{k}_{2}}\rangle_{\lambda_{k}}+\langle{A^{k}_{3}}\rangle_{\lambda_{k}}+\langle{A^{k}_{4}}\rangle_{\lambda_{k}}+\langle{A^{k}_{5}}\rangle_{\lambda_{k}}-\langle{A^{k}_{6}}\rangle_{\lambda_{k}}+\langle{A^{k}_{7}}\rangle_{\lambda_{k}}+\langle{A^{k}_{8}}\rangle_{\lambda_{k}}\\\nonumber&&+\langle{A^{k}_{9}}\rangle_{\lambda_{k}}+\langle{A^{k}_{10}}\rangle_{\lambda_{k}}+\langle{A^{k}_{11}}\rangle_{\lambda_{k}}+\langle{A^{k}_{12}}\rangle_{\lambda_{k}}-\langle{A^{k}_{13}}\rangle_{\lambda_{k}}-\langle{A^{k}_{14}}\rangle_{\lambda_{k}}-\langle{A^{k}_{15}}\rangle_{\lambda_{k}}-\langle{A^{k}_{16}}\rangle_{\lambda_{k}}\bigg| 
	\end{eqnarray}

	Similarly, we can  factorize $|I^{n}_{5,2}|$, $|I^{n}_{5,3}|$, $|I^{n}_{5,4}|$ and $|I^{n}_{5,5}|,$ as  the above equation.	Then using the inequality (\ref{Tavakoli}), for $m=5$, we get 
		\begin{equation}
		\label{A2delta25}
		(\Delta^{n}_{5})_{nl}\leq\prod\limits_{k=1}^{n}\bigg[\sum\limits_{i=1}^{5}\sum\limits_{x_{k}=1}^{16}(-1)^{y^{x_{k}}_{i}} A_{x_{k}}^{k}\hspace{1pt}\bigg]^\fn=\prod\limits_{k=1}^{n}(\eta_5^k)^\fn
		\end{equation}

	  Using the fact that each observable is dichotomic and bounded by the conditions Eq. (\ref{condition5}), we get that  \begin{equation}\eta^k_5=\sum\limits_{i=1}^{5}\bigg|\sum\limits_{x_{k}=1}^{16}(-1)^{y^{x_{k}}_{i}} A_{x_{k}}^{k}\bigg|\leq 16 ,(k\in[n])\end{equation}
	  Putting these in Eq.(\ref{A2delta25}), we obtain $(\Delta^{n}_{5})_{pnc}\leq 16$ as claimed in Eq. (\ref{Delta5}).
	
	Next, we derive the optimal quantum value of $(\Delta^{n}_{5})_{Q}$ using SOS approach. Following the technique developed in Sec. \ref{III} of this paper (for $m=3$), we show that there is a positive semidefinite operator $\langle\gamma^{n}_{5}\rangle\geq 0$, that can be expressed as $\langle \gamma^n_{5}\rangle_{Q}=-(\Delta^{n}_{5})_{Q}+\beta^{n}_{5}$. Here $\beta^{n}_{5}$ is the optimal value can be obtained when $\langle \gamma^n_{5}\rangle_{Q}$ is equal to zero. This can be proven by considering a set of suitable positive operators $M^{n}_{5,i}$ which is polynomial functions of   $A^{k}_{x_{k}}, (\forall k\in[n])$ and $B_{i}$  so that 
	\begin{equation}
	\label{gamma5}
	\langle\gamma^{n}_{5}\rangle=\sum\limits_{i=1}^{5}\dfrac{(\omega_{5,i}^{n})^\fn}{2}|M^{n}_{5,i})|\psi\rangle|^2\end{equation}
	where $\omega^{n}_{5,i}$ is suitable positive numbers and $\omega^{n}_{5,i}=\prod\limits_{k=1}^n(\omega^{n}_{5,i})_{A_{k}}$. The optimal quantum value of $(\Delta^{n}_{5})_{Q}$ is obtained if $\langle \gamma^{n}_{5}\rangle_{Q}=0$, implying that	$|M^{n}_{5,i}|\psi\rangle|=0$ (for notational convenience, we use $|\psi\rangle_{A_1A_2\cdots A_nB}=|\psi\rangle).$\\

	We choose a  suitable set of  positive operators $M^{n}_{5,i}$ are given by
\ba\label{mn5i}
|M^{n}_{5,i}|\psi\rangle|=\prod\limits_{k=1}^n\bigg|\frac{1}{(\omega_{5,i}^{n})_{A_{k}}}\sum\limits_{x_{k}=1}^{16} (-1)^{y^{x_{k}}_i}  A^{k}_{x_{k}}|\psi\rangle
	\bigg|^\fn -|B_{i}|\psi\rangle|^\fn
	\ea	
  Following the argument stated in Sec \ref{III}, we can show that $|M^{n}_{5,i}|\psi\rangle|$ is indeed a positive number.
	Using these in Eq. (\ref{gamma5}), we get 
	$\langle\gamma^{n}_{5}\rangle_{Q}=-(\Delta^{n}_{5})_{Q} +\sum\limits_{i=1}^5(\omega^{n}_{5,i})^\fn$. 	The optimal value of $(\Delta^{n}_{5})_{Q}$ is obtained if  $\langle \gamma^n_{5}\rangle_{Q}=0$.	Hence,
	\begin{eqnarray}(\Delta^{n}_{5})_{Q}^{opt} =\sum\limits_{i=1}^{5}(\omega^{n}_{5,i})^\fn
	\end{eqnarray}
	where
	$(\omega^{n}_{5,i})_{A_{k}}=||\sum\limits_{x_{k}=1}^{16} (-1)^{y^{x_{k}}_i}  A^{k}_{x_{k}}||_{2}$.
	Since  $\omega^{n}_{5,i}=\prod\limits_{k=1}^n(\omega^{n}_{5,i})_{A_{k}}$, by using the  inequality (\ref{Tavakoli}),
	we get \ba\label{wn5i}\sum\limits_{i=1}^{5}(\omega^{n}_{5,i})^\fn\leq \prod\limits_{k=1}^n\bigg(\sum\limits_{i=1}^{5}(\omega^{n}_{5,i})_{A_{k}}\bigg)^\fn\ea
	Using convex inequality, $\sum\limits_{i=1}^{5}(\omega^{n}_{5,i})_{A_{k}}\leq \sqrt{5\sum\limits_{i=1}^{5}\bigg((\omega^{n}_{5,i})_{A_{k}}\bigg)^2}$,  $ k\in[n]$ and noting that each  observable $A_{x_k}^k$ is dichotomic, 
		\begin{eqnarray}
		\label{w25ik}
		\bigg(\sum\limits_{i=1}^{5}\bigg((\omega^{n}_{5,i})_{A_{k}}\bigg)^2&=& \langle\psi|(80+\delta_5)|\psi\rangle
		\end{eqnarray}
		Following the similar method as stated in Appendix A, let $\delta_5 =\sum\limits_{l=1}^{11}(\delta_5)_l$ and $(\delta_5)_l=16-	\langle|\psi_{l}|\psi_{l}\rangle$ and   \begin{eqnarray}
		|\psi_{l}\rangle=\bigg(\sum\limits_{x_{k}=1}^{16} (-1)^{s_l.y^{x_{k}}}A^{k}_{x_{k}}\bigg)|\psi\rangle
		\end{eqnarray}  such that $|\psi\rangle\neq \phi$.
		Since, there are total $11$ elements in the  set $\mathcal{L}_5$, so for different element (say $s_l$) of the set $\mathcal{L}_5$, we will get different $|\psi_{l}\rangle, l\in[11] $.   Then, for each $l$, we get
		\begin{eqnarray}
		\label{5l}
		    \langle\psi_{l}|\psi_{l}\rangle=\langle\psi|(16+(\delta_5)_l)|\psi\rangle \implies \langle(\delta_5)_l\rangle=16-\langle\psi_{l}|\psi_{l}\rangle
		    \end{eqnarray} such that $\sum\limits_{l=1}^{11}\langle(\delta_5)_l\rangle=\langle\delta_5\rangle$.
	Clearly, adding Eqs. of (\ref{5l}), we get $\langle\delta_5\rangle=176-\sum\limits_{l=1}^{11}\langle(\delta_5)_l\rangle$. Hence,  $\langle\delta_5\rangle_{max}$ is obtained when $\langle\psi_{l}|\psi_{l}\rangle=0$, $ \forall l \in[11]$. Since, $|\psi\rangle\neq \phi$, then 	\begin{eqnarray}\label{pnc5}
	\sum\limits_{x_{k}=1}^{16} (-1)^{s_l.y^{x_{k}}}A^{k}_{x_{k}}=0
		\end{eqnarray}
	Hence, to obtain the optimal value of  $(\Delta^{n}_{5})_{Q}^{opt}$, the  conditions given in Eq.(\ref{pnc5}) have to be satisfied. Using them , $\langle\delta_5\rangle_{max}=176$. Thus from Eq. (\ref{w25ik}), we get $\sum\limits_{i=1}^{5}\bigg((\omega^{2}_{5,i})_{A_{k}}\bigg)^2=256$. Plugging it in the above mentioned inequality (\ref{wn5i}), finally we get $\sum\limits_{i=1}^{5}(\omega^{n}_{5,i})^\fn\leq \prod\limits_{k=1}^{n}\sum\limits_{i=1}^{5}(\omega^{n}_{5,i})_{A_{k}}^\fn\leq16\sqrt{5}$ i.e.,  \ba(\Delta^{n}_{5})_{Q}^{opt}= 16\sqrt{5} \ea 
	
	A straightforward but lengthy  algebraic calculations using the  conditions that is derived in Eq.(\ref{pnc5}), and following the similar procedure of solving  Eqs. (\ref{AK3}-\ref{c3d}), we get that   Alice's observables has to satisfy the following relations to obtain the optimal quantum value. 
	$\{A^k_1,A^k_j\}=\frac{6}{5}, j=2,\ldots ,6$,  	$\{A^k_1,A^k_j\}=\frac{2}{5}, j=7,\ldots ,16$. $\{A^k_2,A^k_j\}=\frac{2}{5}, j=3,\ldots ,6$,  	$\{A^k_2,A^k_j\}=\frac{6}{5},  j=7,8,10,13$,  $\{A^k_2,A^k_j\}=-\frac{2}{5}, j=9,11,12,13,14,15,16$, 
	$\{A^k3,A^k_j\}=\frac{2}{5},  j=4,5,6$, $\{A^k_3,A^k_j\}=\frac{6}{5}, j=7,9,11,14$, $\{A^k_3,A^k_j\}=-\frac{2}{5}, j=8,10,12,13,15,16$, 
	$\{A^k_4,A^k_j\}=\frac{2}{5}, j=5,6$,  $\{A^k_4,A^k_j\}=\frac{6}{5}, j=8,9,12,15$, $\{A^k_4,A^k_j\}=-\frac{2}{5}, j=7,10,11,13,14,16$,  $\{A^k_5,A^k_j\}=\frac{2}{5}, j=5,6$, $\{A^k_5,A^k_j\}=\frac{6}{5}, j=10,11,12,16$, $\{A^k_5,A^k_j\}=-\frac{2}{5}, j=7,8,9,13,14,15$,
	 $\{A^k_6,A^k_j\}=\frac{6}{5}, j=13,14,15,16$, $\{A^k_4,A^k_j\}=-\frac{2}{5}, j=7,8,9,10,11,12$, $\{A^k_7,A^k_j\}=\frac{2}{5}, j=8,9,10,11,13,14$, $\{A^k_7,A^k_j\}=-\frac{6}{5}, j=12,15,16$,	$\{A^k_8,A^k_j\}=\frac{2}{5}, j=9,10,12,13,15$, $\{A^k_8,A^k_j\}=-\frac{6}{5}, j=11,14,16$, $\{A^k_9,A^k_j\}=\frac{2}{5}, j=11,12,14,15$, $\{A^k_9,A^k_j\}=-\frac{6}{5}, j=10,13,16$, $\{A^k_{10},A^k_j\}=\frac{2}{5}, j=11,12,13,16$, $\{A^k_{10},A^k_j\}=-\frac{6}{5}, j=14,16$, $\{A^k_{11},A^k_j\}=\frac{2}{5}, j=12,14,16$, $\{A^k_{11},A^k_j\}=-\frac{6}{5}, j=13,15$, $\{A^k_{12},A^k_j\}=\frac{2}{5}, j=15,16$, $\{A^k_{12},A^k_j\}=-\frac{6}{5}, j=13,14$, $\{A^k_{13},A^k_j\}=\frac{2}{5}, j=14,15,16$, $\{A^k_{14},A^k_j\}=\frac{2}{5}, j=15,16$, $\{A^k_{15},A^k_{16}\}=\frac{2}{5}$.  \\

	Now, one can find Bob's observables from the optimization conditions i.e.,  for each $i\in[5]$, 	$M^{n}_{5,i}|\psi\rangle=0$. This in turn fixes the required state which has to be at least two copies of  two-qubit maximally entangled state shared by each Alice$_k$ and Bob. 
\end{widetext}

\end{document}